\documentclass[notitlepage]{paper}
\usepackage{epsfig}
\begin{document}
\begin{center}
\LARGE \textbf{Pitch Tracking of Acoustic Signals based on Average Squared Mean Difference Function}
\end{center}
\begin{center}
\textbf{Roudra Chakraborty}
\\ \footnotesize{Jadavpur University, Kolkata.}
\\ \footnotesize{roudra@gmail.com}
\\ 
\textbf{Debapriya Sengupta}
\\ \footnotesize{Indian Statistical Institute, Kolkata.}
\\ \footnotesize{dps@isical.ac.in}
\\
\textbf{Sagnik Sinha}
\\ \footnotesize{Jadavpur University, Kolkata.}
\\ \footnotesize{sagnik62@yahoo.co.in}
\end{center}

\centering\mbox{\textbf{\small{Abstract}}}
\begin{quotation}
\textbf{In this paper, a method of pitch tracking based on variance minimization of locally periodic subsamples of an acoustic signal is presented. Replicates along the length of the periodically sampled data of the signal vector are taken and locally averaged sample variances are minimized to estimate the fundamental frequency. Using this method, pitch tracking of any text independent voiced signal is possible for different speakers.}
\end{quotation}

\centering\mbox{\large{I. INTRODUCTION}}
\begin{quotation}

Extraction or determination of fundamental frequency (or pitch) of a speech signal is a fundamental problem in both speech processing and speaker recognition. The typical pitch range for a male human being is 80-200 Hz, and for females 150-350 Hz. Many methods to extract the pitch of speech signals have been proposed. Improvements in accuracy of performance, robustness against noise of these methods are still desired. As a whole, we do not have any reliable and accurate method for pitch extraction. Also measuring the period of a speech waveform, varying in and with the detailed structure of the waveform, can be quite difficult. Another problem is automatic selection of the window of the voiced speech segments.

Autocorrelation method [Rabiner, 10] and the average magnitude difference function (AMDF) method [Ross et. al., 15] are known to be the most primitive standard methods to find pitch. Based on these two methods several refinements like auditory modeling [Cosi et. al., 3], probabilistic AMDF modeling [Jamieson et. al., 5], real-time digital hardware pitch detector [Rabiner et. al., 12], semiautomatic pitch detector (SAPD) [Rabiner et. al., 13], automatic formant analysis [Rabiner et. al., 14], weighted autocorrelation [Shimamura et. al., 16], modified autocorrelation and AMDF [Tan et. al., 17], projection measure technique [Yuo et. al., 18], pseudo-pitch synchronous analysis [Zilca et. al., 19] and many more [Rabiner et. al., 11] are proposed. Some other ideas on pitch extraction have also been discussed in the paper [Marchand, 9] and some tutorials [Gerhard, 4] and [Campbell, 6].

For an idealized speech signal in a stationary noisy environment the following mathematical abstraction has been consistently assumed in this paper.
\begin{equation}
y_n=\sum_{j}a_j\exp\left\{\mbox{i}f(j)\,n\right\} + \sum_{j}c_{j}z_{j}\exp\left\{\mbox{i}jn\right\},
\end{equation}
where $\left\{z_j\right\}$ is an uncorrelated sequence of random variables or white noise and $\{c_j\}$ are coefficients of discrete spectrum of stationary noise. In case the main signal $\left\{y_n\right\}$ possesses a pure fundamental frequency (or pitch),
which is again another idealized view, it is assumed in this paper that $f(j) = jf_p$ for a suitable pitch value $f_p$. The actual situation can become complicated further if the idealized signal contains multiple pitch streams or is convoluted with a channel filter. Therefore, from a practical point of view,  the estimation of the pitch of a signal is essentially a statistical problem. Here we propose a new method for extraction of fundamental frequency of speech signal in clean and to some extent noisy environment using simple statistical techniques. Motivationally, the technique has similarity with Zero Crossing based techniques (Kedem, [8] and Gerhard, [4]), however, the theory is much simpler and statistical in nature. The main novelty of our approach actually lies in formulating the problem in this manner and putting a certain number of standard measures (such as, Autocorrelation and AMDF) of pitch in the same framework with our proposed measure (Average Squared Mean Difference Function or ASMDF). This way a more comprehensive approach is presented and left open for further refinements. 

The remainder of this paper is organized as follows. Section II describes the motivation and the principle of the proposed method. In Section III, we show the results of preliminary tests for the proposed method and comparing with some standard pitch detection methods we confirm the effectiveness of our method. In Section IV, we give an error analysis of the method. In section V, we conclude this paper giving views regarding further development that can be done. In the appendix, given as in Section VI, important calculations, which show the link between the proposed method and the autocorrelation function, are explained with a short note on computational complexity of the proposed method.

\end{quotation}
\newpage
\centering\mbox{\large{II. PROPOSED METHOD}}
\begin{quotation}

Given a discrete time signal $y=(y_1, y_2, \ldots, y_n)$ (which is the real part of the complex idealized signal (1)), the autocovariance function is defined as \begin{equation}r_y(k)=\frac{1}{n-\left|k\right|}\sum_{i=0}^{n-\left|k\right|}(y_{i}-\overline{y})(y_{i+k}-\overline{y}),\end{equation} and the autocorrelation function is defined by \begin{equation}\rho_y(k)=\frac{r_y(k)}{r_y(0)}\end{equation} defined for all $n$ and lag $k$.

A variation of autocorrelation analysis for measuring the periodicity of voiced speech uses the average magnitude difference function (AMDF), defined by the relation $$D_y(k)=\frac{1}{n-\left|k\right|}\sum_{j=1}^{n-\left|k\right|}\left|y_{j+k}-y_{j}\right|.$$

In case of AMDF, $D_y(k)$ has been approximated by a scalar multiple of $$\frac{1}{n-\left|k\right|}\left\{\sum_{j=1}^{n-\left|k\right|}(y_{j+k}-y_{j})^2\right\}^{1/2}$$ which again has been approximated by the scalar multiple of $\left[2\left\{r_y(0)-r_y(k)\right\}\right]^{1/2},$ where $r_y(k)$ is the autocovariance defined as above. These approximations may suppress calculations important for pitch extraction.

Also AMDF can be considered as a replica of Gini's mean difference formula ([Goon et. al., 7], page 233-234). But the new method described below gives more optimal results as it is a replica of variance of the data set. So we propose a new function which has been linked to the autocorrelation and refined in the Appendix. 

Consider the voiced segment $y=(y_1,y_2,\ldots, y_n)$ in a digital speech signal. Since most speech signals can be viewed as a quasi-periodic sequence the fundamental frequency may not be uniquely defined mathematically. In our approach we estimate the fundamental frequency by statistically enhancing the most significant harmonics present in $y$. 

We describe below our algorithm for estimation of the pitch in this voiced segment. For $1 \leq i \leq n$, and $k \geq 1$ we consider the downsampled subsets (windows) of the original signal,
$$ y_{i,k} = ( y_{i+pk}: p=0, \pm{1}, \pm{2}, \ldots ).$$
Note that due to finiteness of the data stream, for each $(i,k)$ pair we have to consider only those values of $p$ so that $y_{i+pk}$ is within the range. Furthermore, for several $(i,k)$ pairs, $y_{i,k}$ will become singleton and they will not come under further considerations. It is worth  bringing in the issue of aliasing here. The parameter $k$ is treated like a trial wavelength parameter so that smaller values of $k$ relates to higher frequencies and vice versa. We restrict lower limit of this parameter and assume $k \geq k_0$ as an adjustment for the Nyquist rate based on the original sampling rate of the signal. However, as our goal is estimation of the fundamental frequency of the voiced part we shall generally be interested in the higher range of values of $k$. This raises the issue of aliasing. Again based on the sampling rate of the signal an upperbound for $k$, namely $k_{max}$, needs to be set so that overall sampling rate after downsampling stays above 40 kHz. One of the basic assumption is that the fundamental frequency of the voiced part is estimable (in terms of both Nyquist rate and aliasing) for the given signal. 

Next, let for each $k \geq k_0$, define
$$ S_k = \left\{ i : 1\leq i \leq n; \ y_{i,k} \, \mbox{has at least two elements}\right\},
$$
and, let $q_k$ be the number of elements in $S_k$. This automatically sets an upperbound for $k \leq k_{max} < ([(n+1)/2] -1)$ ([x] being the greatest integer less than x). Finally define for $k$ values with $q_k >0$,
\begin{equation}
g(k)=\frac{1}{q_k} \, \sum_{i \in S_k}\, \mbox{Var}(y_{i,k}),
\end{equation}
where $ \mbox{Var}(y_{i,k})$ denotes the sample variance of the signal values in the subset $y_{i,k}$. It is interesting to note that aliasing, statistical precision of the sample variances (requiring larger values of $q_k$ over a large range of $k$) of downsampled signals and robustness of the method under noisy environments considered are largely related issues.

Let $f_0$ be the sample rate of the original speech signal and $f(k)=\frac{f_0}{k},$ where $k_0 \leq k \leq k_{max}$.
In view of (4), $g$ can be thought of as a function of $f$. Also $g$ can be thought of as a mean squared mutual difference function which can be approximated with the standard autocorrelation function.

Let $i$ be the index of the second minimum of the components of $g(k)$, i.e. $g(i)=min_{k} g(k)$ Then $f_p=f(i)$ is referred as the estimated fundamental frequency of the speech signal of $y$.

We further assume that the white noise sequence $\left\{z_j\right\}$ in (1) is Gaussian and carry out the Likelihood Ratio test of the null hypothesis, $H_0 : E(y_{i,k}) = \mu_{i}(1,1,\ldots)$, the ASMDF statistic $g(k)$ coincides with a measure of departure from this null hypothesis. This is the motivation behind the main proposal in the present paper.

\end{quotation}
\centering\mbox{\large{III. EXPERIMENTS}}

\centering\mbox{A. Synthetic Data}

\begin{quotation}
\textbf{Experiment No. 1.} We have taken the function $y_n=sin(x_n)$ for $x_n=\frac{2n\pi}{55}; n=0, 1, 2, \ldots, 11000.$ Then we have plotted $g(k)$ for the 100th window for different $k$ and also plotted $r_y(k)$ (autocorrelation values) and $D_y(k)$ (AMDF values) against $k$ in the same graph.
\end{quotation}
\begin{quotation}
\begin{tabular}{c|c|c}
\epsfig{file=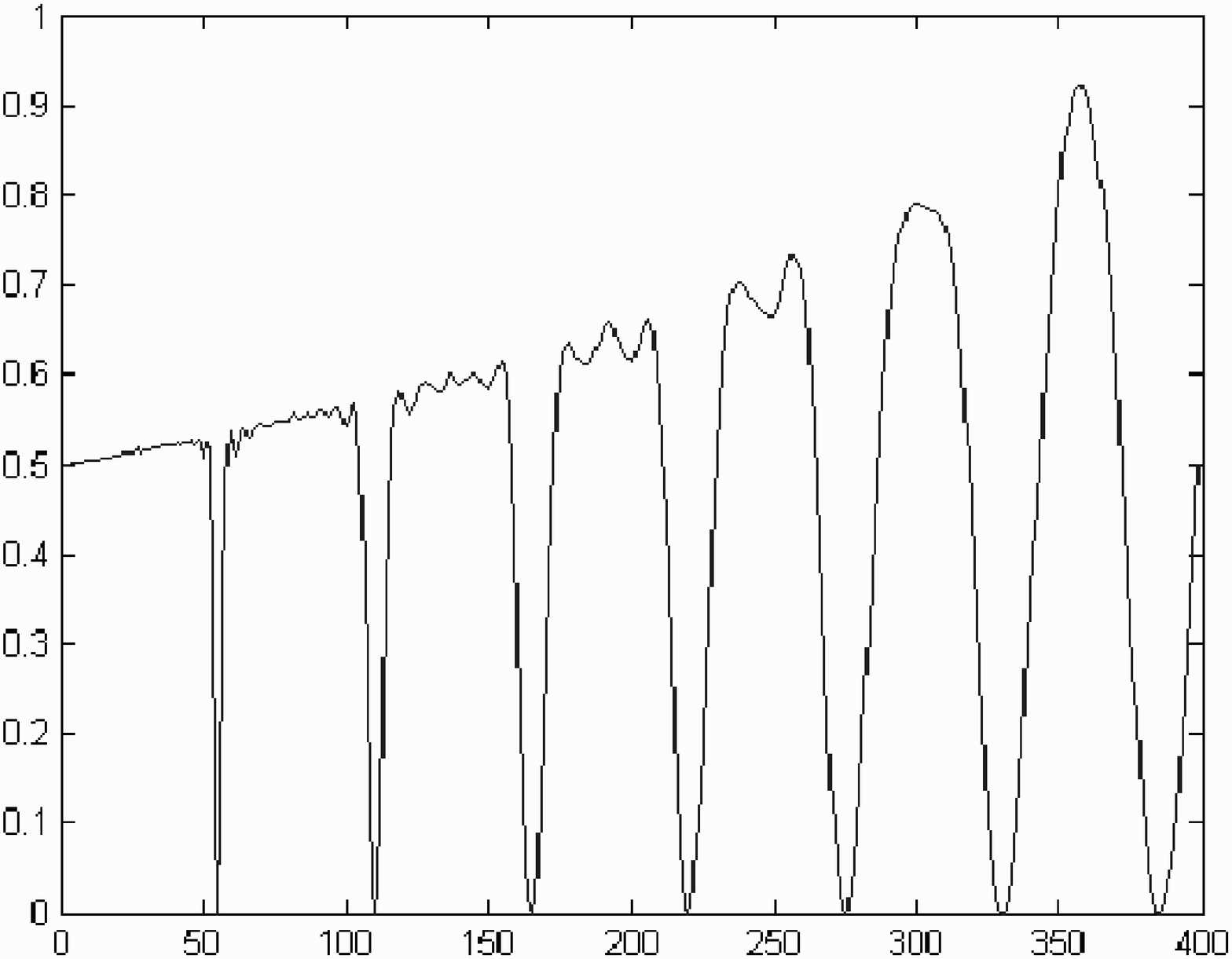,height=4.2cm,width=3.2cm,angle=90} & \epsfig{file=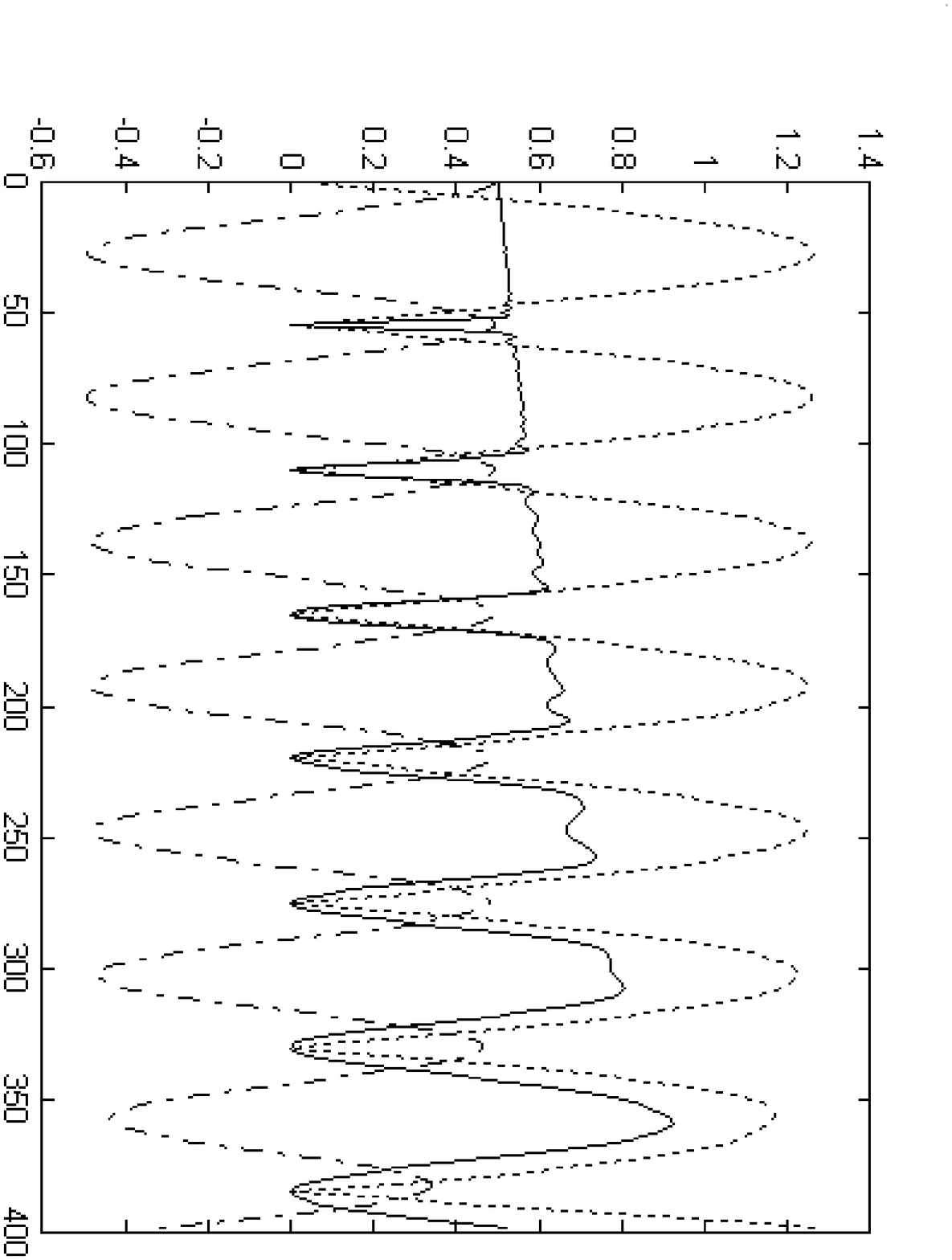,height=4.2cm,width=3.2cm,angle=90} &
\epsfig{file=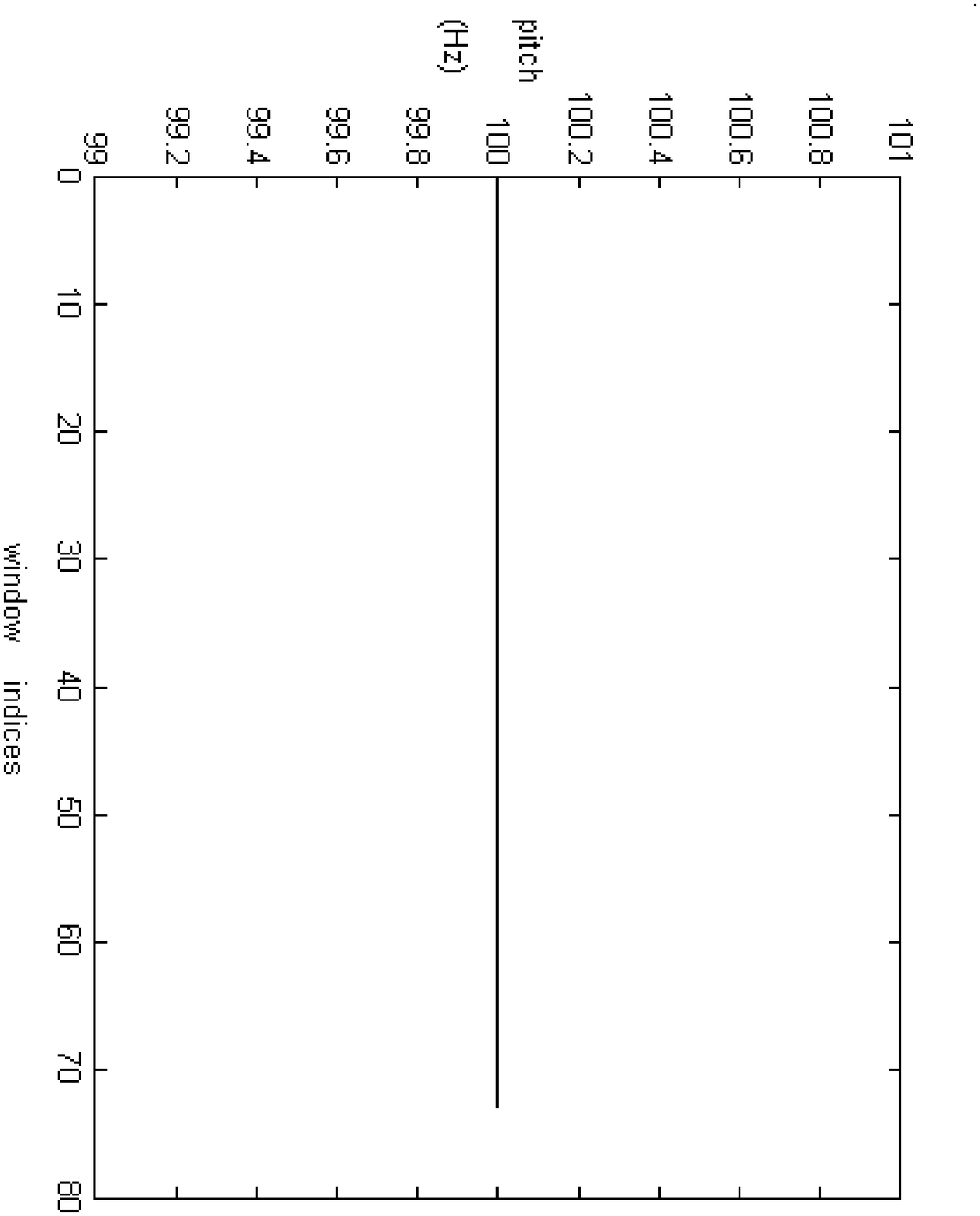,height=4.2cm,width=3.2cm,angle=90} \\
\small Figure 1 : Graph of $g(k)$ & 
\small Figure 2 : Graph of $g(k)$, $r_y(k)$ (dashed) &
\small Figure 3 : Graph of pitch \\
\small against k for $y=sin(x)$ &
\small and $D_y(k)$ (dotted) for $y=sin(x)$ &
\small for y = sin(x) \\
\end{tabular}
\end{quotation}
\begin{quotation}
Here we see that peaks in the autocorrelation graph and dips in the AMDF and ASMDF occur for the same $k$. Next we present the graph of fundamental frequency, which is expected to be 100Hz, for the same data with sample rate 11000Hz, a window size of 400 samples and shifting the window along 55 consecutive samples, using ASMDF. Autocorrelation and AMDF give identical graph.
\end{quotation}
\begin{quotation}
\textbf{Experiment No. 2.} We have taken $a_n=sin(b_n)$ for $b_n=\frac{2n\pi}{55}; n=0, 1, 2, \ldots, 5500$ and $c_n=cos(d_n)$ for $d_n=\frac{5n\pi}{56}; n=0, 1, 2, \ldots, 5500$. Then $y$ has been taken as a linear combination of $a_n$ and $c_n$ as $y=0.47a_n+0.59c_n$. Now we plot $g(k)$ for the 100th window for different $k$ and also plot $r_y(k)$ (autocorrelation values) and $D_y(k)$ (AMDF values) against $k$ in the same graph.
\end{quotation}
\begin{quotation}
\begin{center}
\begin{tabular}{c|c}
\epsfig{file=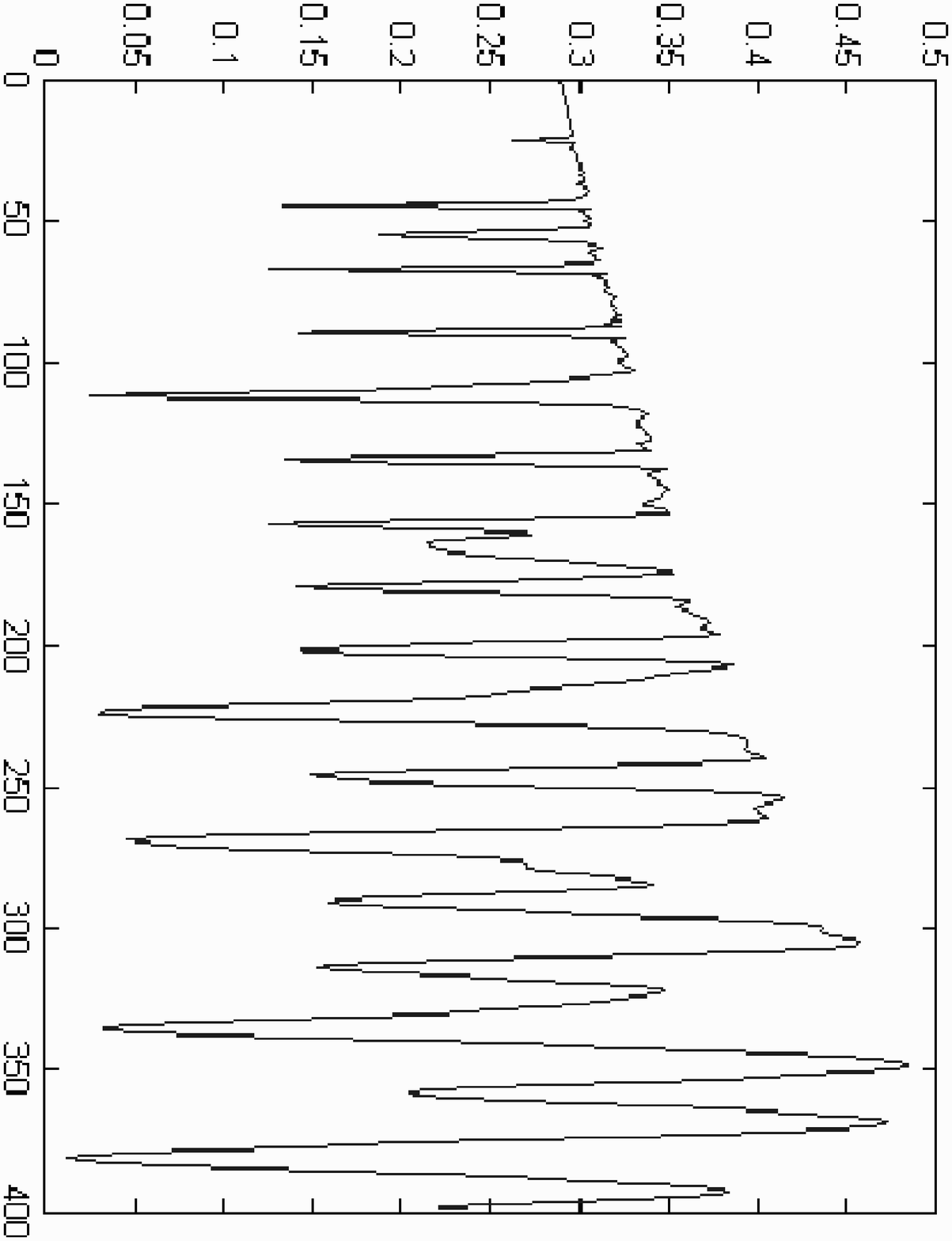,height=5cm,width=3.5cm,angle=90} & \epsfig{file=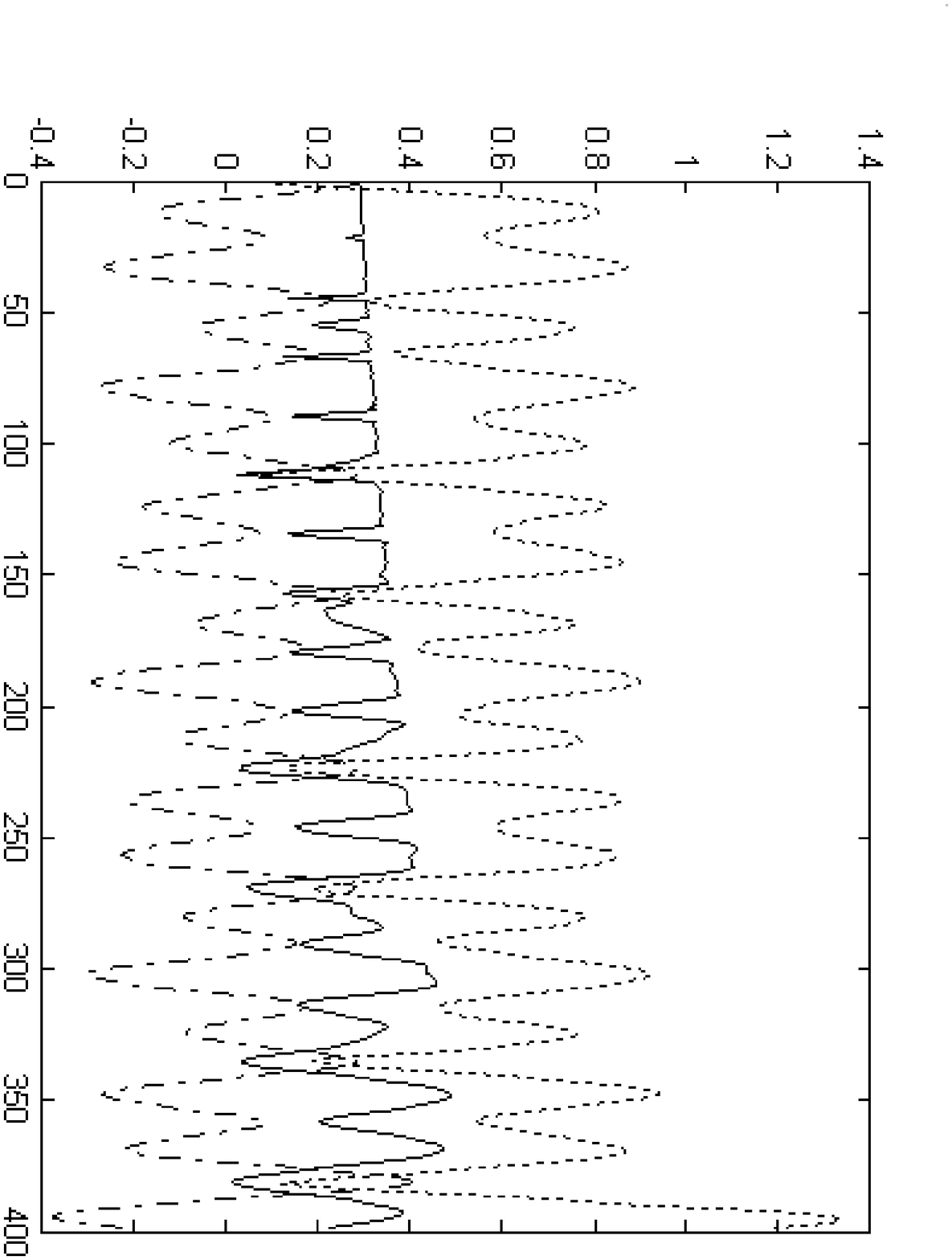,height=6cm,width=4cm,angle=90} \\
\small Figure 4 : Graph of $g(k)$ & 
\small Figure 5 : Graph of $g(k)$, $r_y(k)$ (dashed) \\
\small against k for $y=0.47a_n+0.59c_n$ &
\small and $D_y(k)$ (dotted) against k for $y=0.47a_n+0.59c_n$ \\
\end{tabular}
\end{center}
\end{quotation}
\begin{quotation}
Here we see that peaks in the autocorrelation and dips in AMDF graphs tally with each other but dips in ASMDF sometimes matches the above two and sometimes falls just short of them. The reason is the peaks and the dips for all three methods are supposed to appear at the indices which are multiples (approximately) of the smallest index. Here the first peak and dip are at the 22nd index for all three methods. The second peak appears for autocorrelation at the 46th index and the second dip for AMDF appears there too. But for ASMDF, the second dip appears at the 45th index which is a better approximation to find pitch. Thus the graph of fundamental frequency, which is expected to be 250Hz $(5500/\left\{2\times gcd([55], [56\times 2/5])=250\right\})$ for the same data with sample rate 5500Hz, a window size of 400 samples and shifting the window along 55 consecutive samples, is found the following way
\end{quotation}
\begin{quotation}
\begin{center}
\begin{tabular}{c|c}
\epsfig{file=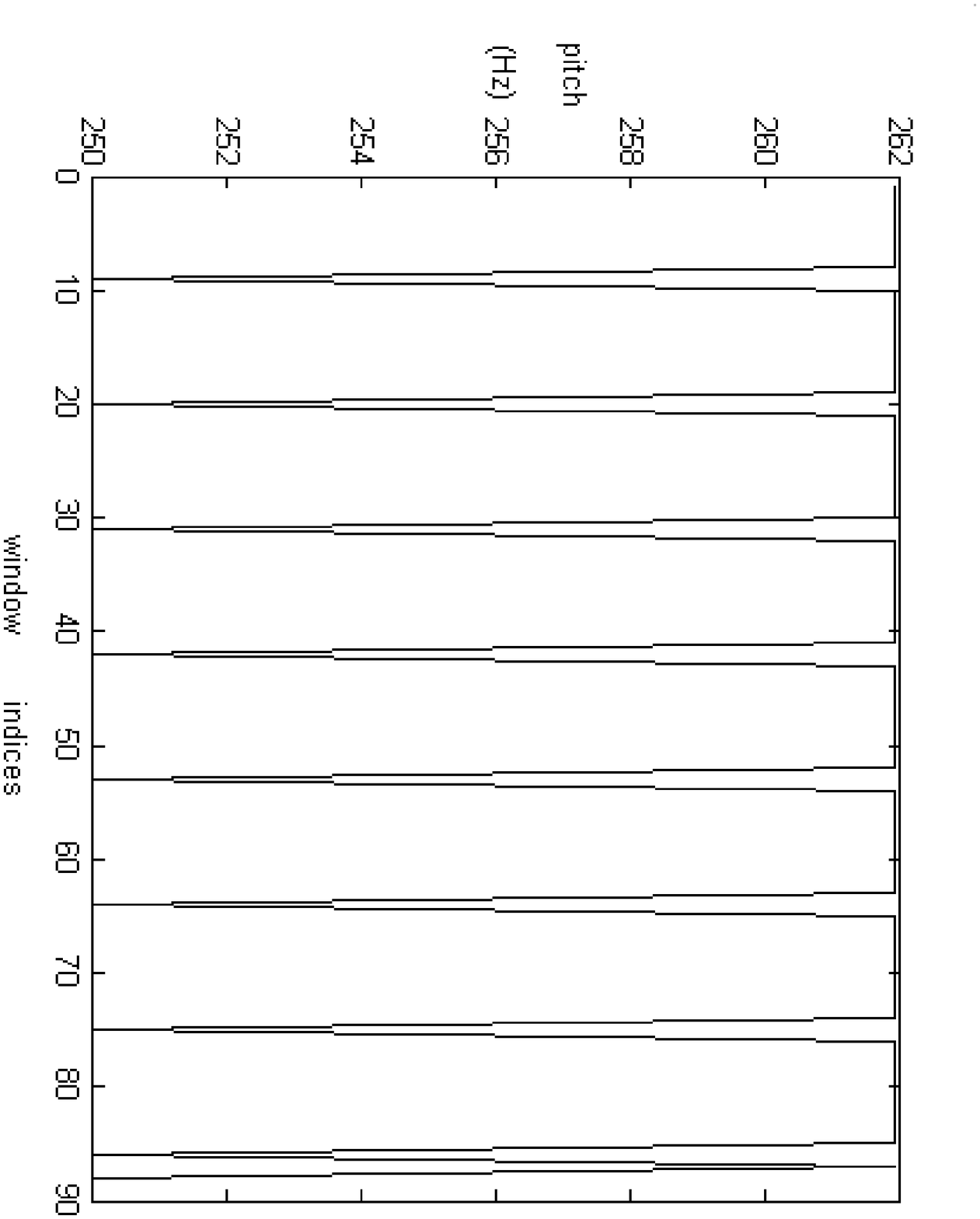,height=6cm,width=4cm,angle=90} & \epsfig{file=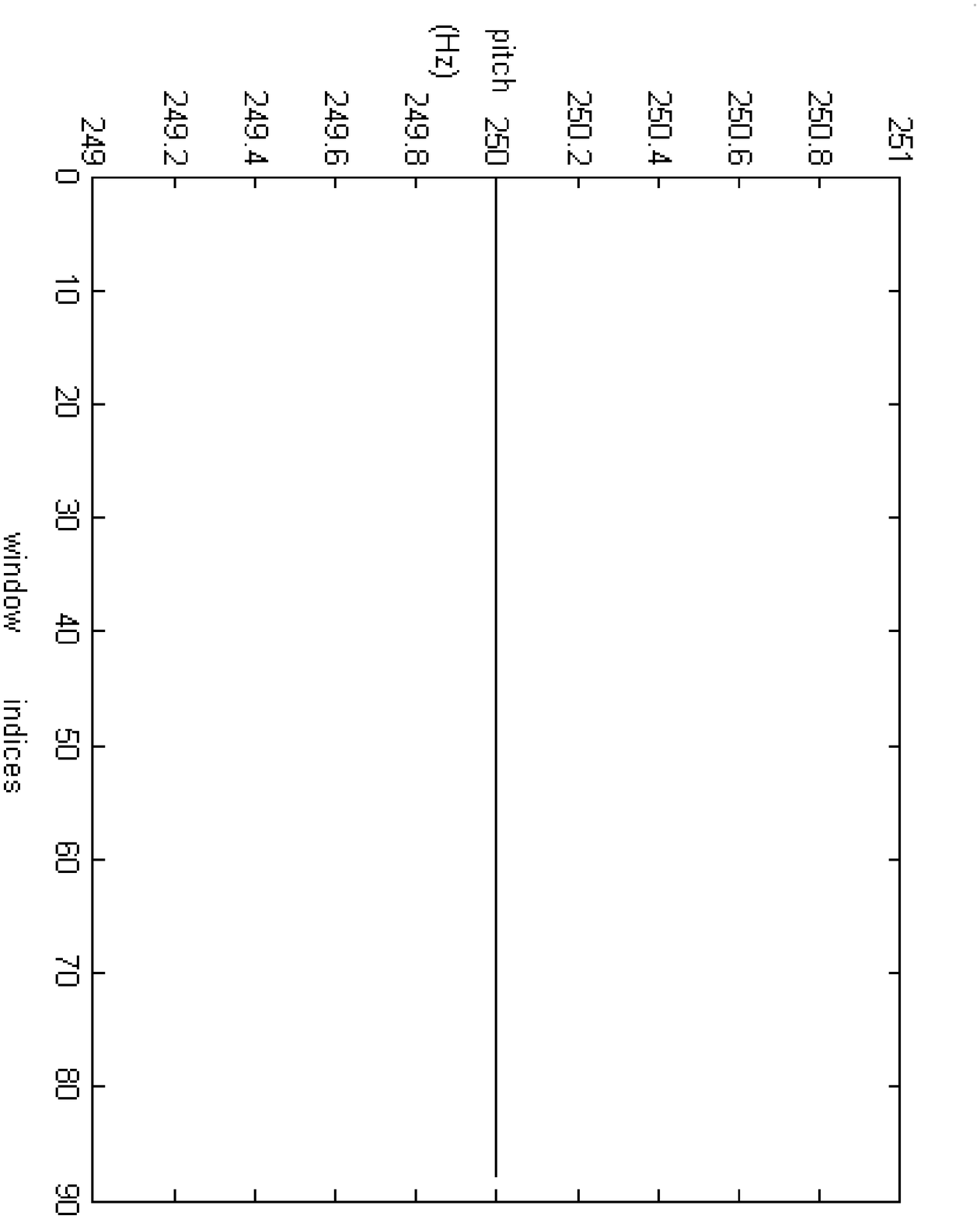,height=6cm,width=4cm,angle=90} \\
\footnotesize Figure 6 : Graph of pitch for AMDF & 
\footnotesize Figure 7 : Graph of pitch for autocorrelation and \\
\footnotesize for $y=0.47a_n+0.59c_n$ &
\footnotesize ASMDF for $y=0.47a_n+0.59c_n$ (since they are identical) \\
\end{tabular}
\end{center}
\end{quotation}
\centering\mbox{B. Speech Data}

\begin{quotation}
\textbf{Experiment No. 3.}
Clean speech signals with unknown true pitch values were obtained from IViE Corpus [1]. Speech samples were uttered by one female speaker (F1) and one male speaker (M1). Each of such speech signals consisted of a maximum of five English words, which were sampled at different rates. Taking window size of 1000 samples we found data sets of pitch and the following graphs.

To investigate the accuracy of the ASMDF, we have conducted experiments which compare it with two conventional methods. They are the methods of Autocorrelation and AMDF.

Autocorrelation method does the correlation analysis frame-by-frame to the estimated average pitch period of the speaker. The property of this function is that $r_y(k)$ is large when $y_n$ has similar value with $y_{n+k}$. If $y_n$ has a pitch period $i$, then $r_y(k)$ has peaks at the integral multiples of $i$. Obviously $r_y(0)$ is maximum among these values, the second largest being $r_y(i)$. Other maxima usually decrease as $k$ increases. Therefore using this method we can estimate $i$ from the location of the peak at $k=i$.

A variation of autocorrelation analysis for measuring the periodicity of voiced speech uses the AMDF. The separation of the nulls that appear in calculating $D_y(k)$ is a direct measure of the pitch period.

\end{quotation}

\begin{quotation}
\begin{center}
\begin{tabular}{c|c|c}
\epsfig{file=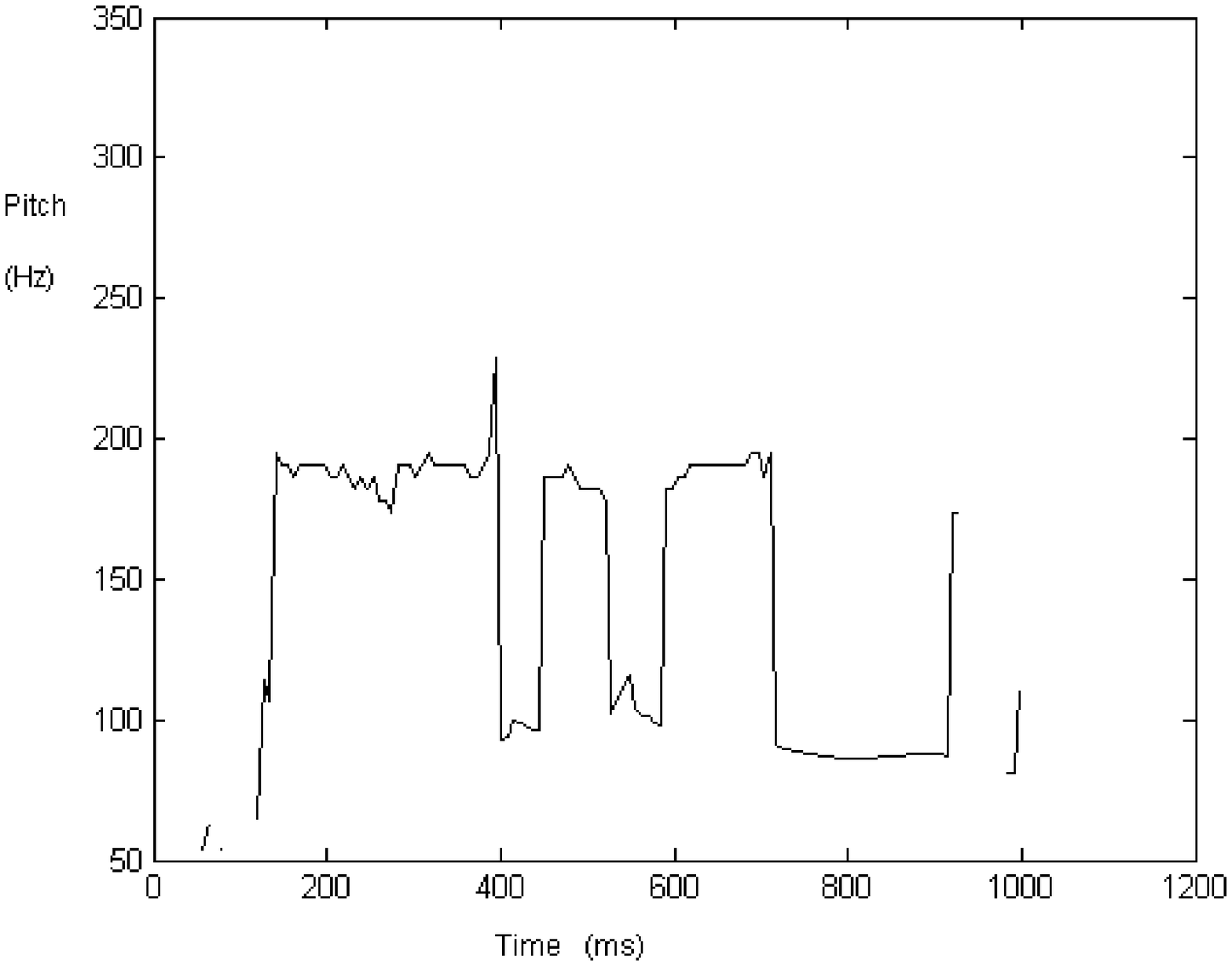,height=4cm,width=6cm} & 
\epsfig{file=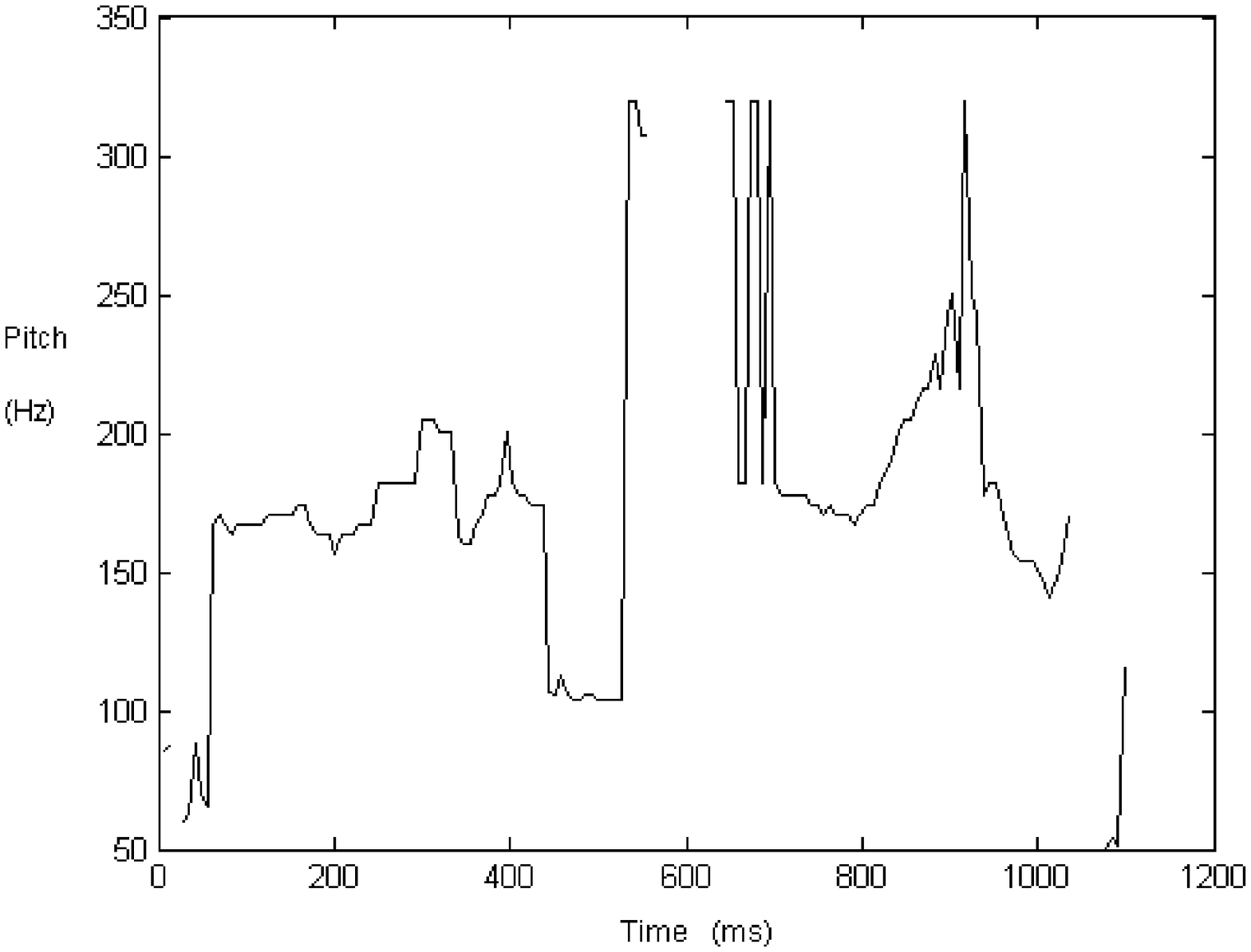,height=4cm,width=6cm} \\
\scriptsize Figure 8 (a) : Graph for pitch of & 
\scriptsize Figure 9 (a) : Graph for pitch of \\
\scriptsize speaker F1 using ASMDF &
\scriptsize speaker M1 using ASMDF \\

\epsfig{file=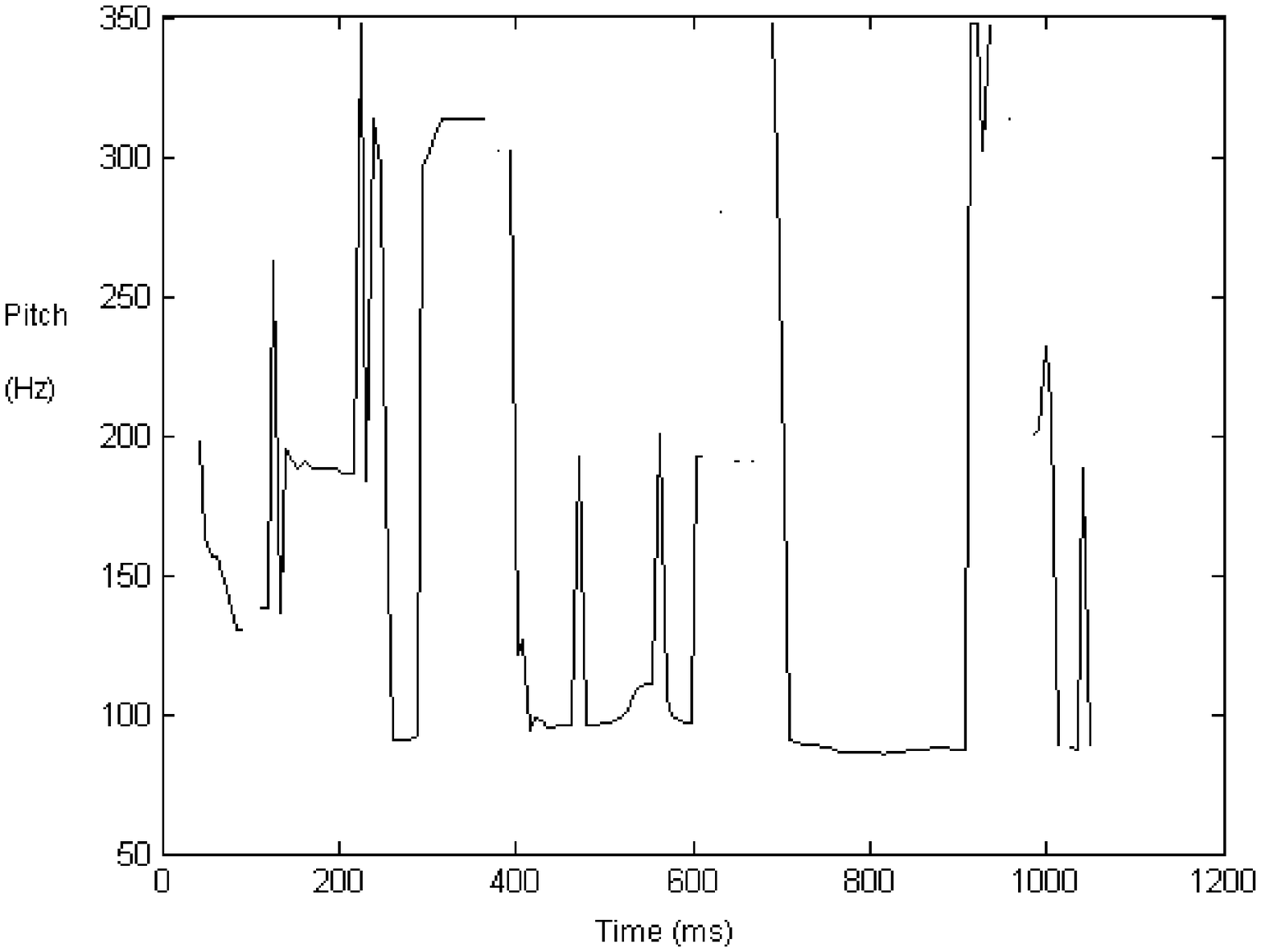,height=4cm,width=6cm} &
\epsfig{file=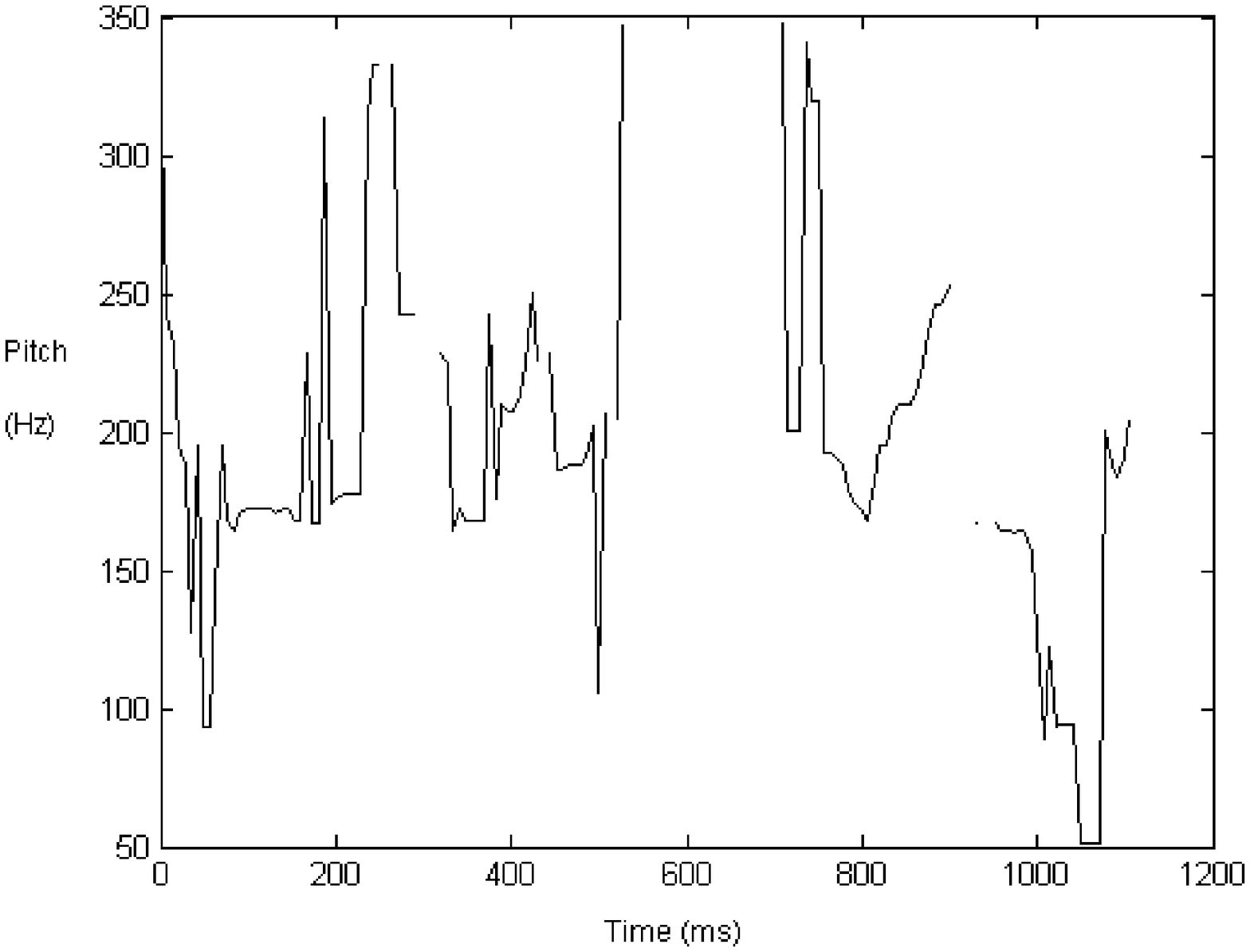,height=4cm,width=6cm} \\
\scriptsize Figure 8 (b) : Graph for pitch of & 
\scriptsize Figure 9 (b) : Graph for pitch of \\
\scriptsize speaker F1 using AMDF &
\scriptsize speaker M1 using AMDF \\

\epsfig{file=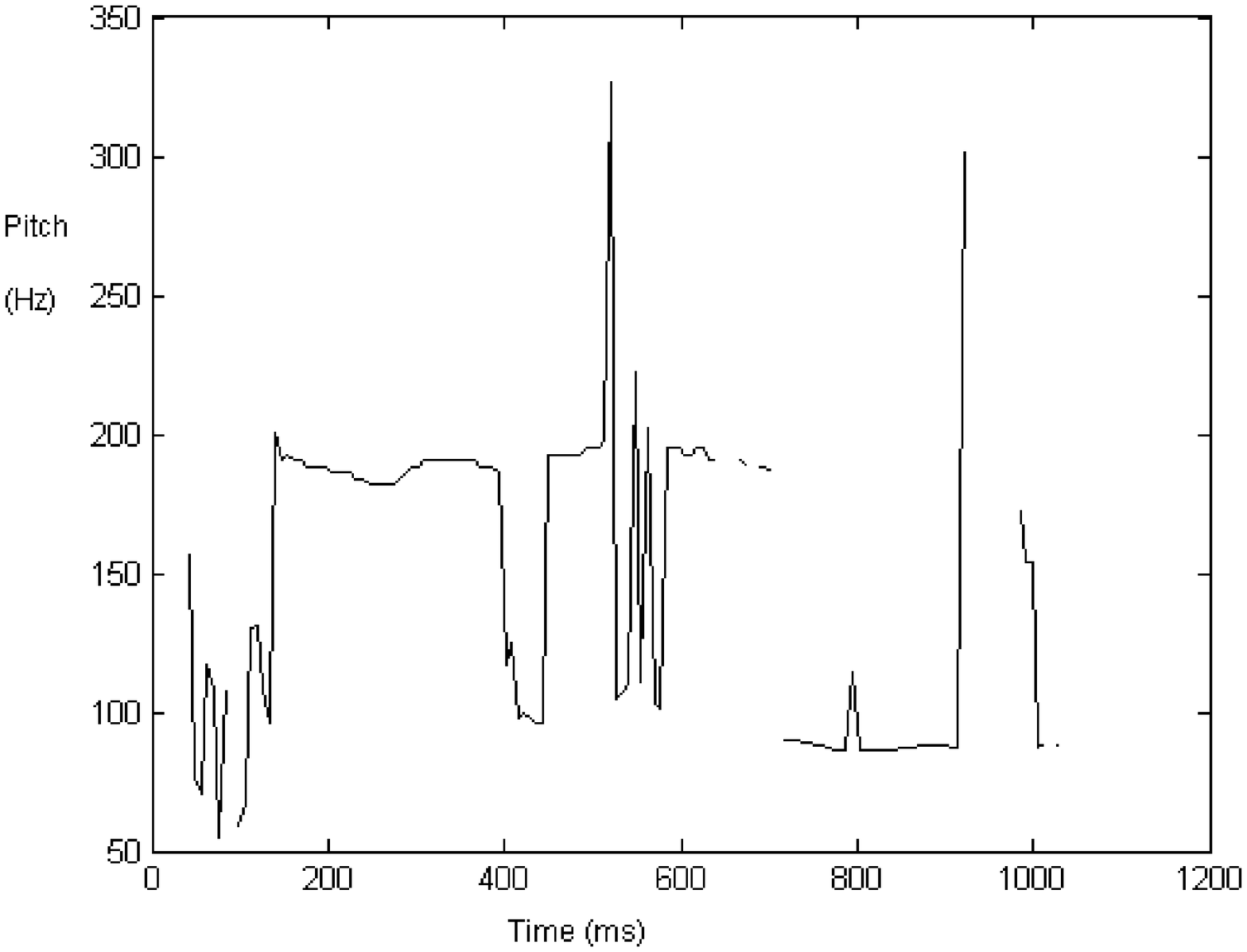,height=4cm,width=6cm} &
\epsfig{file=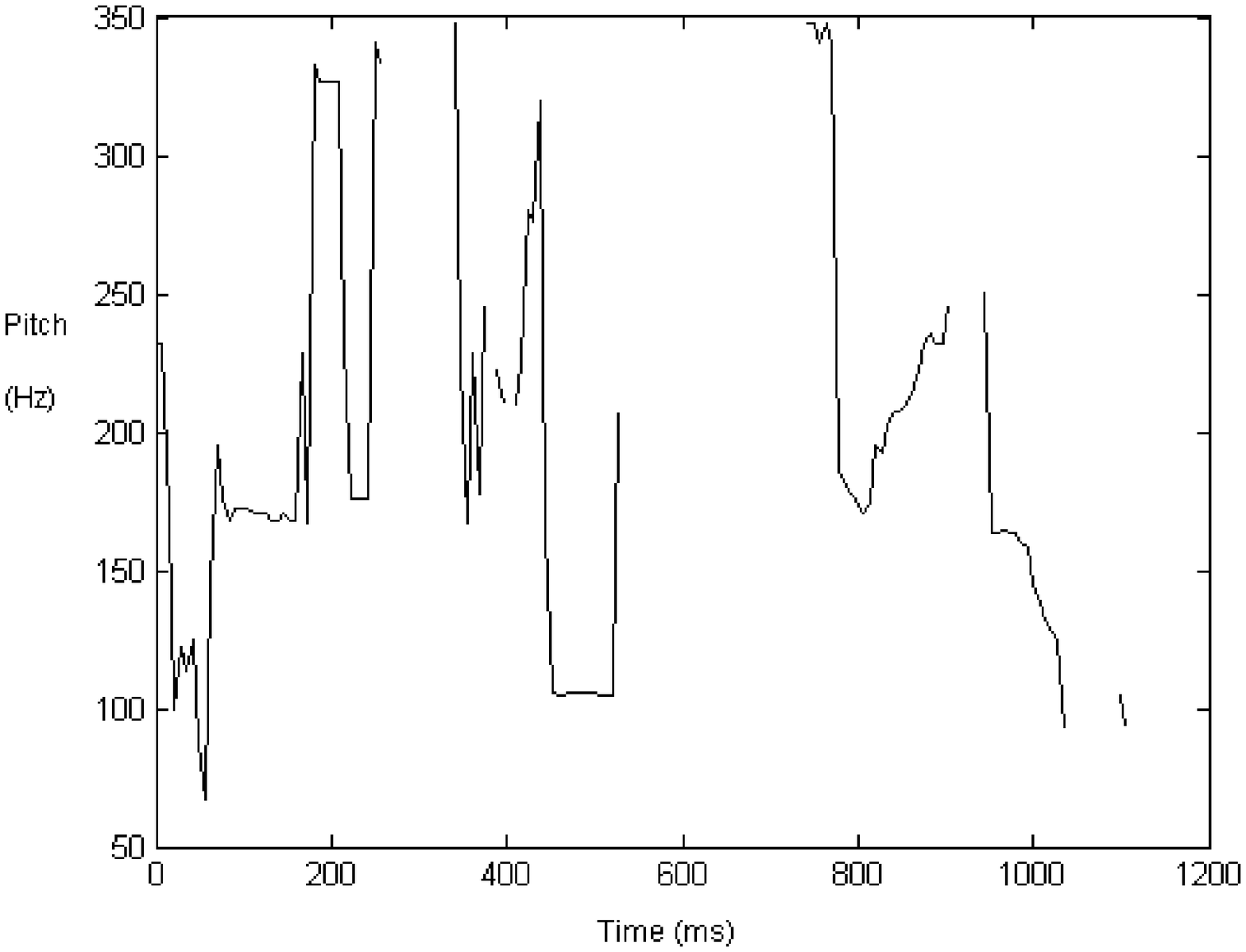,height=4cm,width=6cm} \\
\scriptsize Figure 8 (c) : Graph for pitch of & 
\scriptsize Figure 9 (c) : Graph for pitch of \\
\scriptsize speaker F1 using Autocorrelation &
\scriptsize speaker M1 using Autocorrelation \\
\end{tabular}
\end{center}
\end{quotation}

\begin{quotation}

Observations :
F1 : Here major fluctuations around 280th, 470th and 920th ms for the method autocorrelation and AMDF are observed, whereas no major fluctuations but minor ones are there in the segment for the method ASMDF.

M1 : Here the method ASMDF is much consistent with the methods autocorrelation and AMDF. From 180th and 230th ms major fluctuations are there for the methods autocorrelation and AMDF but not for ASMDF.
\end{quotation}
\begin{quotation}
\textbf{Experiment No. 4.}
One male (RL) and one female speaker (SB) each spoke 50 sentences, out of which, fifteen speeches (with known pitch or Laryngeal frequency contour in XMG format) were taken from FDA Evaluation Database [2] for experiment. Taking window size of 400 samples, we found data sets of pitch and the following graphs (where sample rate of each the Laryngograph waveform is given to be 20000Hz) of the first (001) speeches. All other graphs were observed to have similar behaviour. All the graphs have time (in ms) as horizontal axis and pitch (in Hz) as vertical axis. 

Graphs of pitch of speaker RL speaking speech 001:
\end{quotation}
\begin{quotation}
\begin{center}
\begin{tabular}{c|c}
\epsfig{file=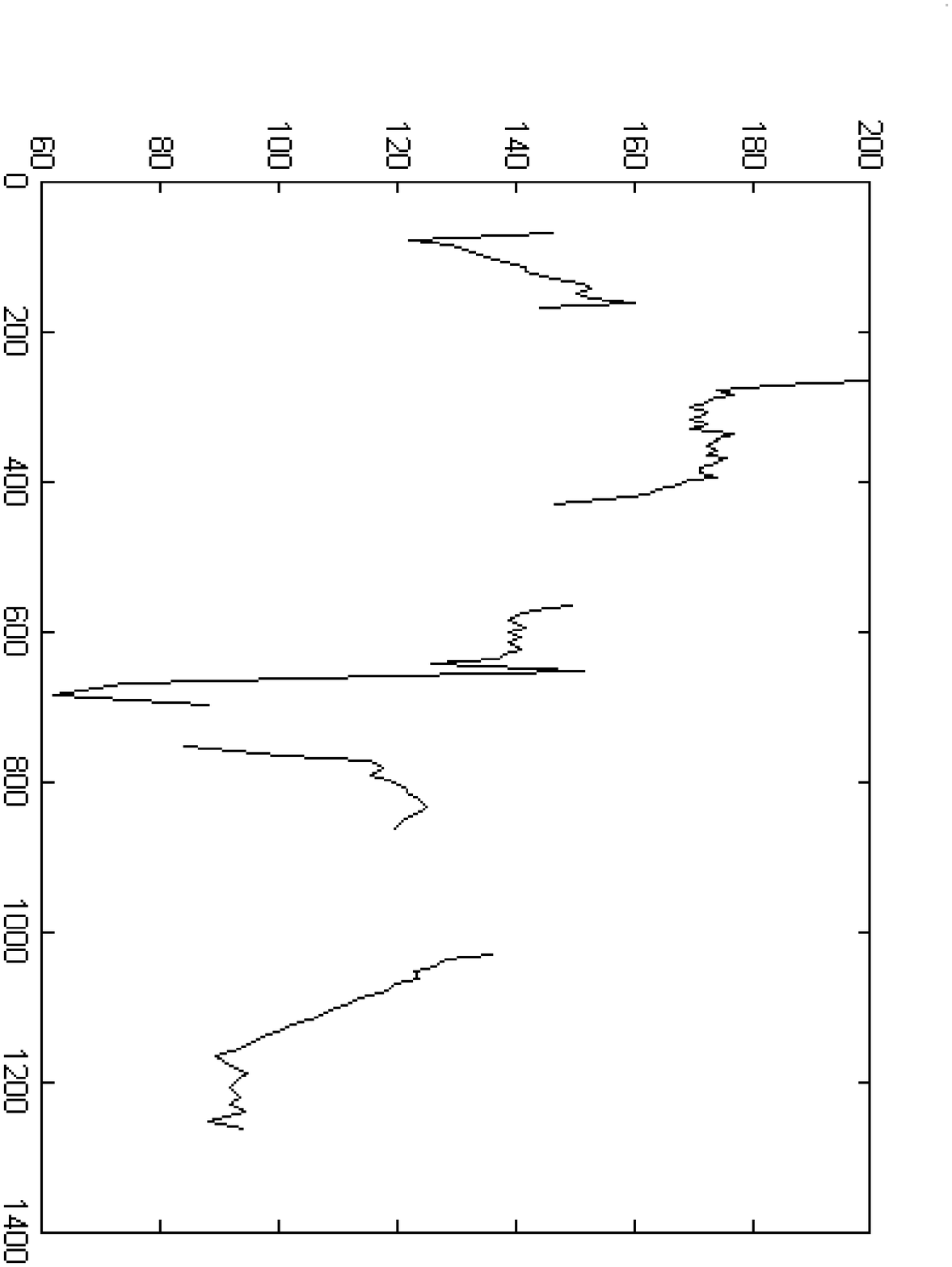,height=6cm,width=4cm,angle=90} & \epsfig{file=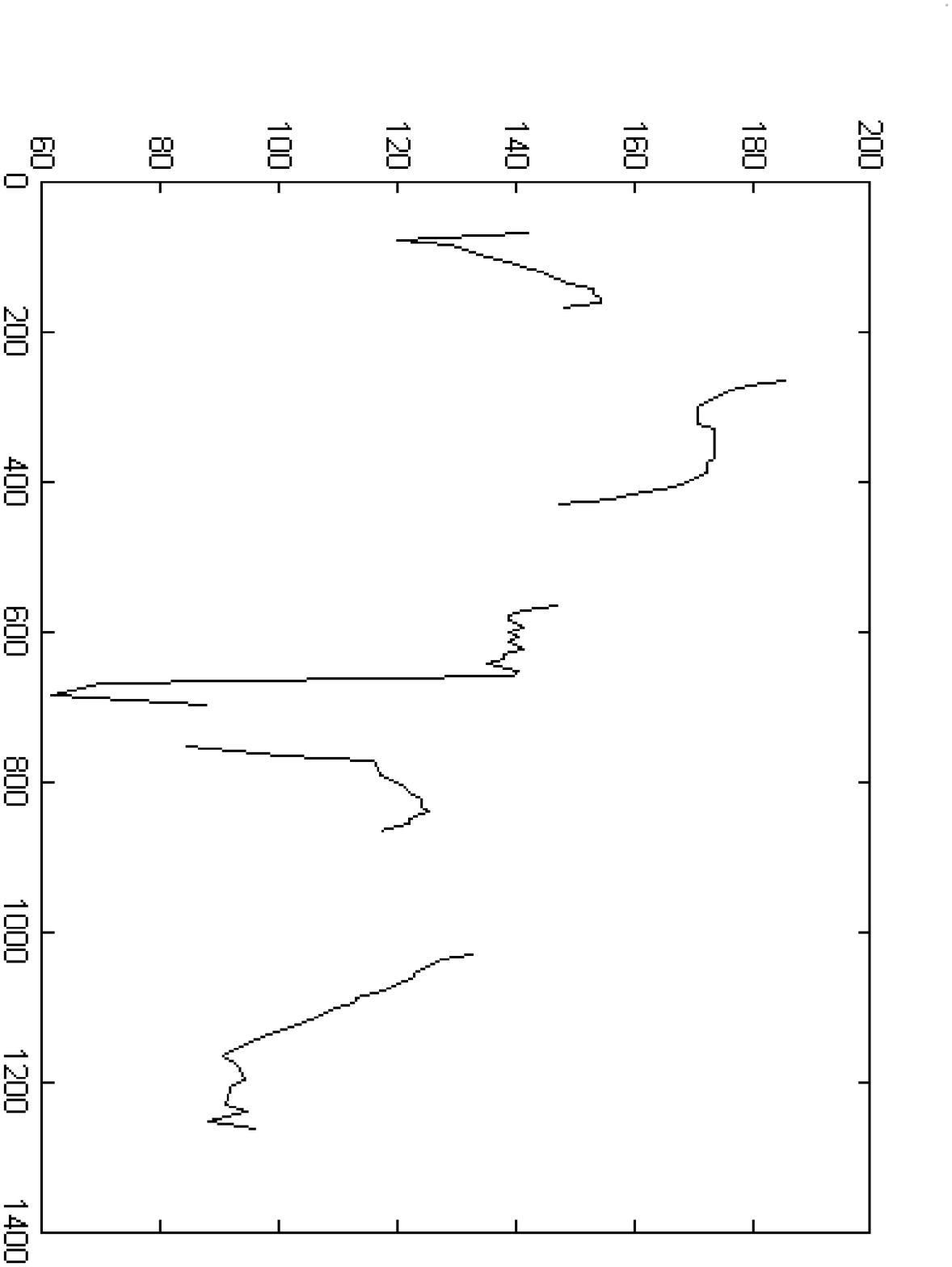,height=6cm,width=4cm,angle=90} \\
\end{tabular}
\end{center}
\footnotesize 
\begin{center}
Fig. 10 (a), (b): Graphs of true and ASMDF pitch values respectively.
\end{center}
\end{quotation}
\begin{quotation}
\begin{center}
\begin{tabular}{c|c}
\epsfig{file=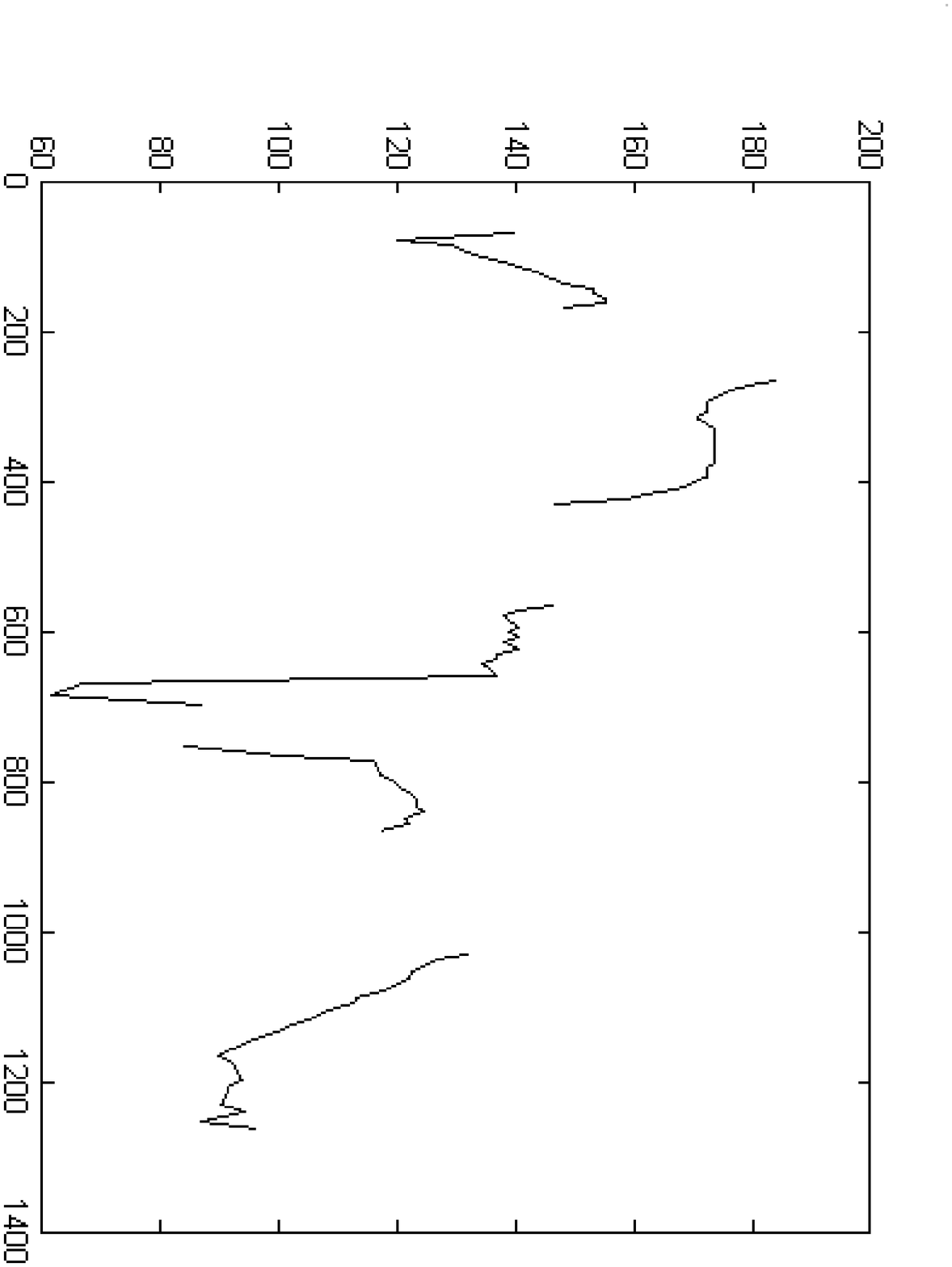,height=6cm,width=4cm,angle=90} & \epsfig{file=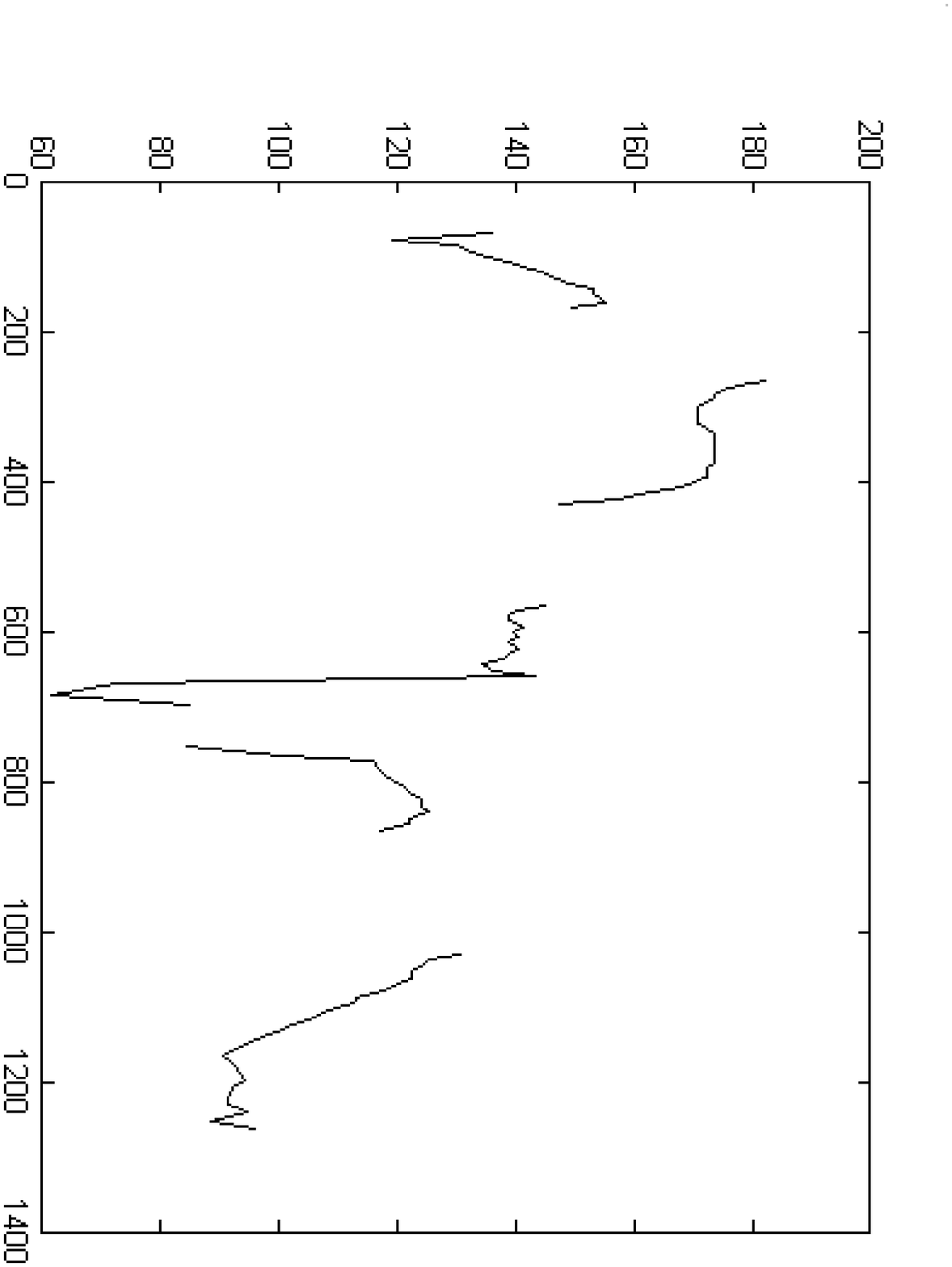,height=6cm,width=4cm,angle=90} \\
\end{tabular}
\end{center}
\footnotesize 
\begin{center}
Fig. 10 (c), (d): Graphs of Autocorrelation and AMDF pitch values respectively.
\end{center}
\end{quotation}
\begin{quotation}

Graphs of pitch of speaker SB speaking speech 001:
\end{quotation}
\begin{quotation}
\begin{center}
\begin{tabular}{c|c}
\epsfig{file=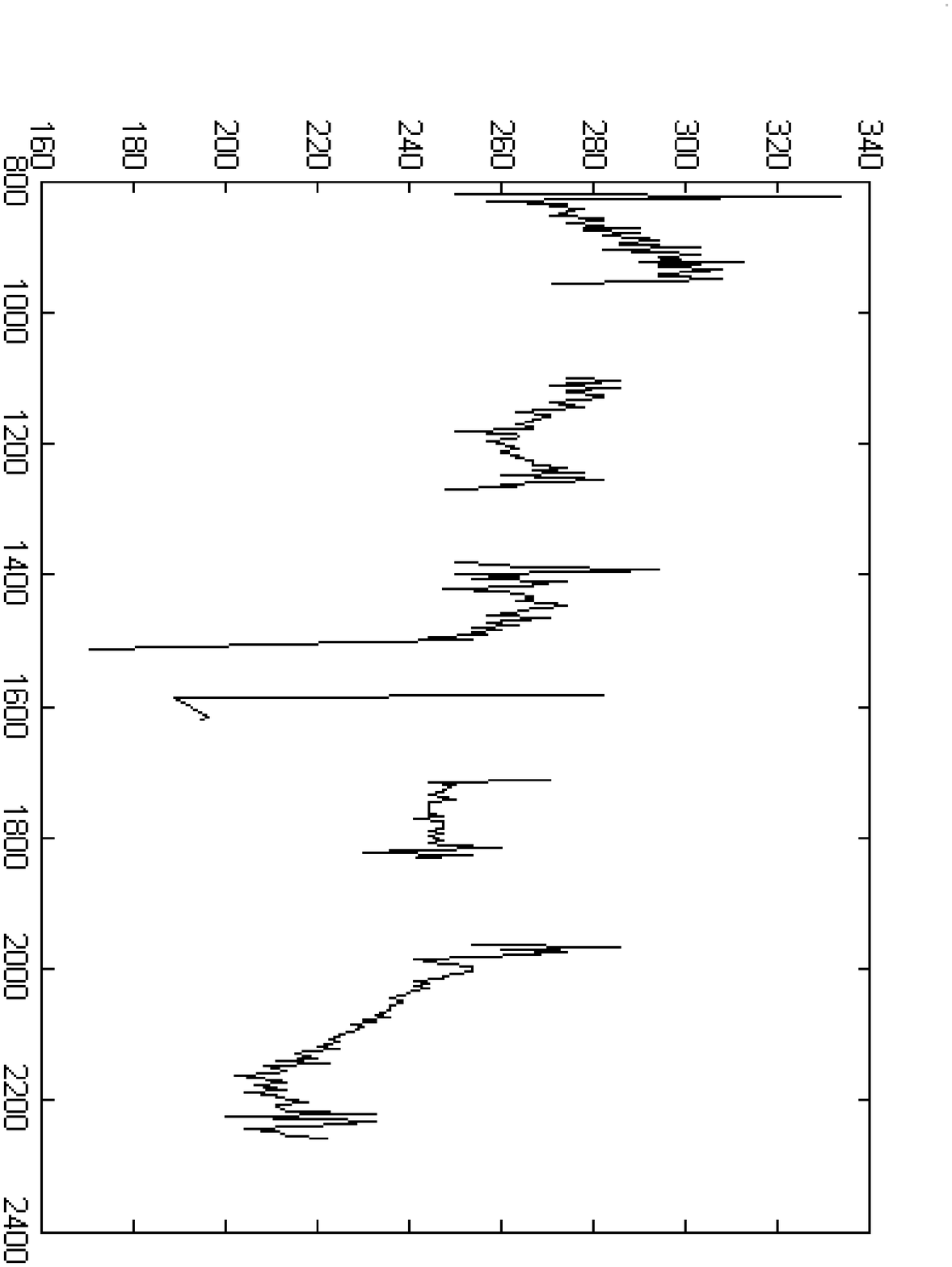,height=6cm,width=4cm,angle=90} & \epsfig{file=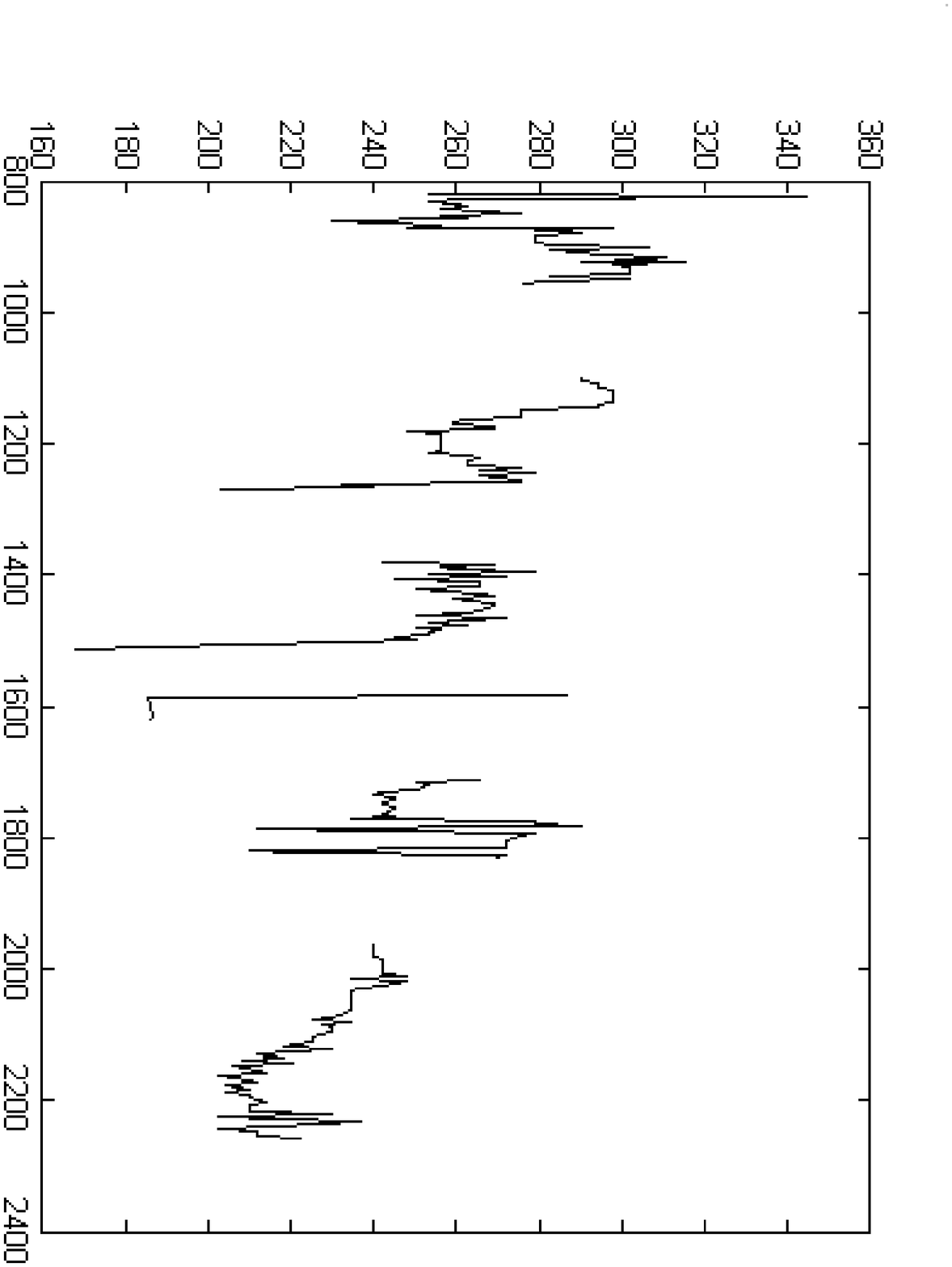,height=6cm,width=4cm,angle=90} \\
\end{tabular}
\end{center}
\footnotesize 
\begin{center}
Fig. 11 (a), (b): Graphs of true and ASMDF pitch values respectively.
\end{center}
\end{quotation}
\newpage
\begin{quotation}
\begin{center}
\begin{tabular}{c|c}
\epsfig{file=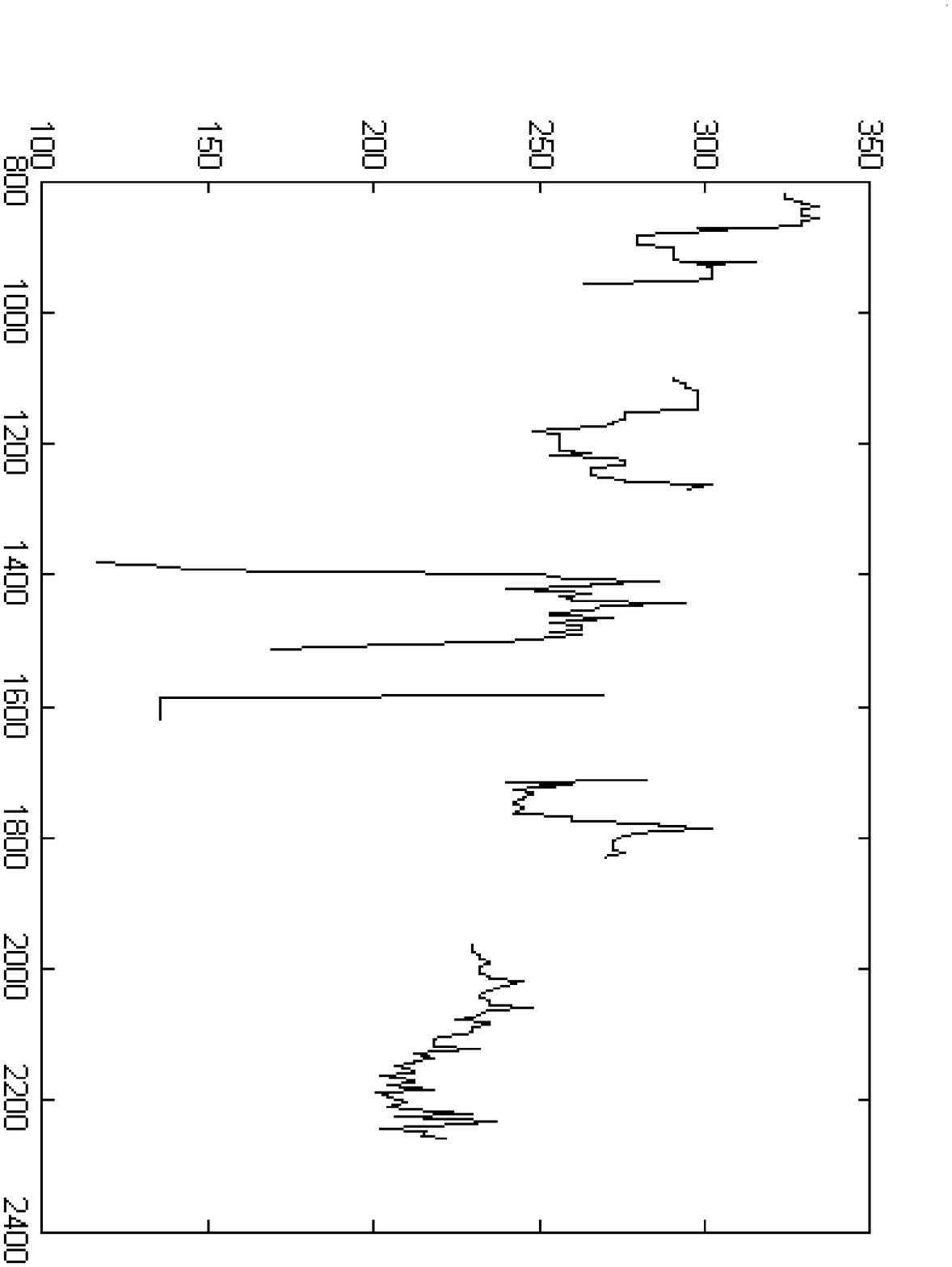,height=6cm,width=4cm,angle=90} & \epsfig{file=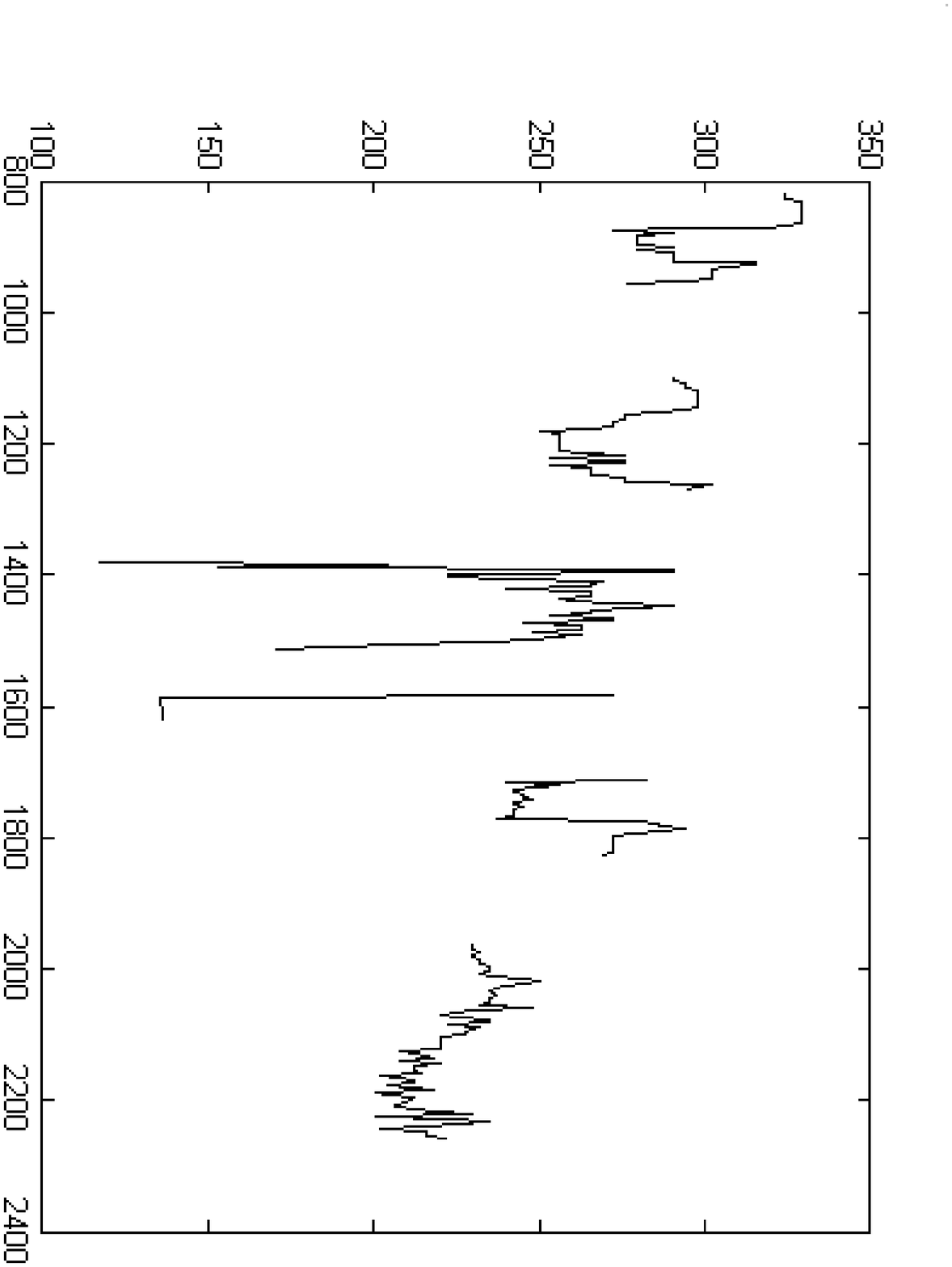,height=6cm,width=4cm,angle=90} \\
\end{tabular}
\end{center}
\footnotesize 
\begin{center}
Fig. 11 (c), (d): Graphs of Autocorrelation and AMDF pitch values respectively.
\end{center}
\end{quotation}

\centering\mbox{C. Synthetic and Noisy Data (without true pitch values)}

\begin{quotation}
\textbf{Experiment No. 5.}
Tones of seven notes, each of length 1.6 seconds (approx.), have been played by a synthesizer of model no. SA21 from Casio in its harmonica mode and recorded in Acer Travelmate 240 Laptop using Audacity 1.2.3 with sample rate 44100 Hz. Following pitch graphs (with pitch axis against k axis) of the note ``Do'' have been found using the three methods. All other graphs from the remaining notes were observed with similar behaviour.
\end{quotation}
\begin{quotation}
\begin{flushleft}
\begin{tabular}{c|c|c}
\epsfig{file=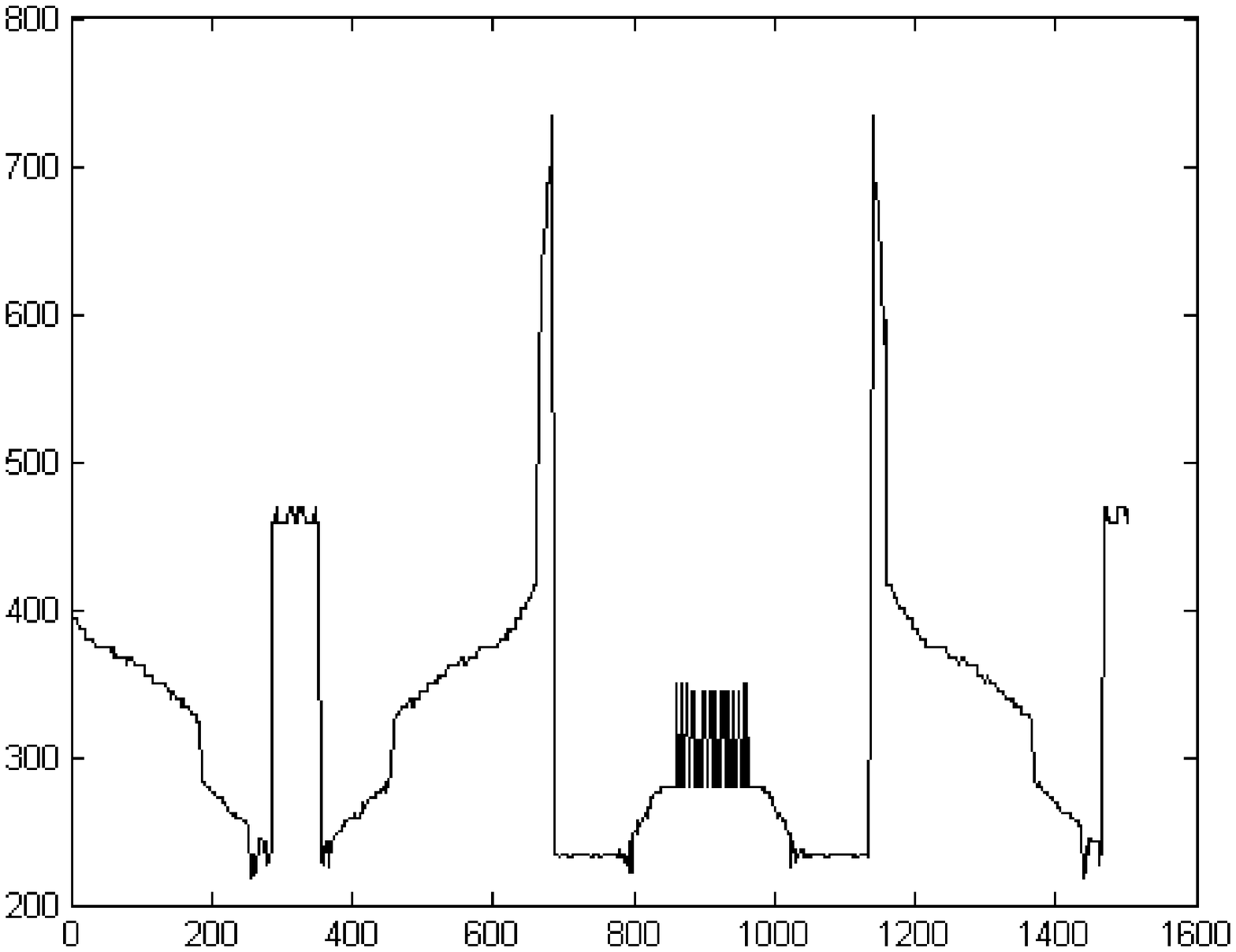,height=3.2cm,width=4.8cm} & \epsfig{file=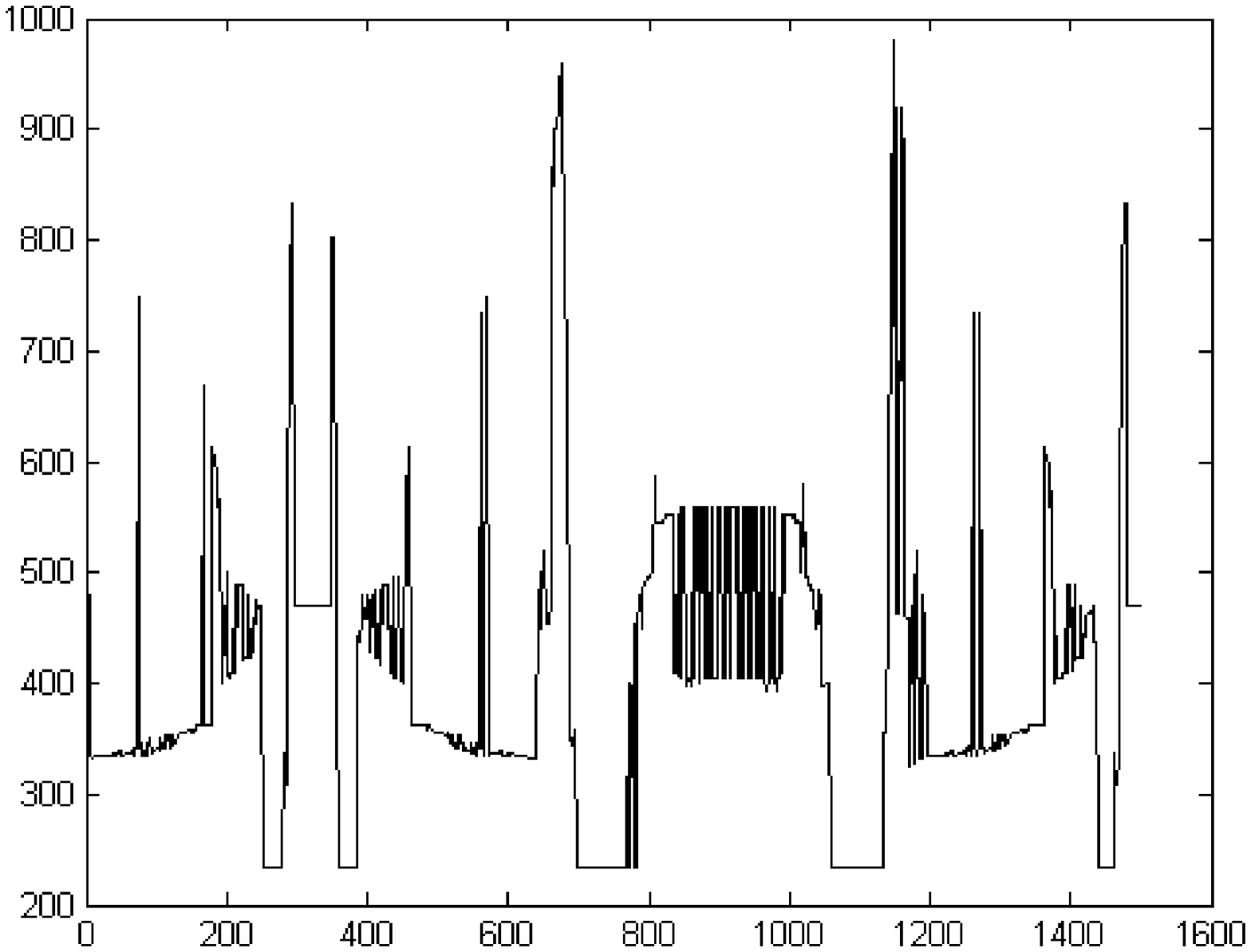,height=3.2cm,width=4.8cm} &
\epsfig{file=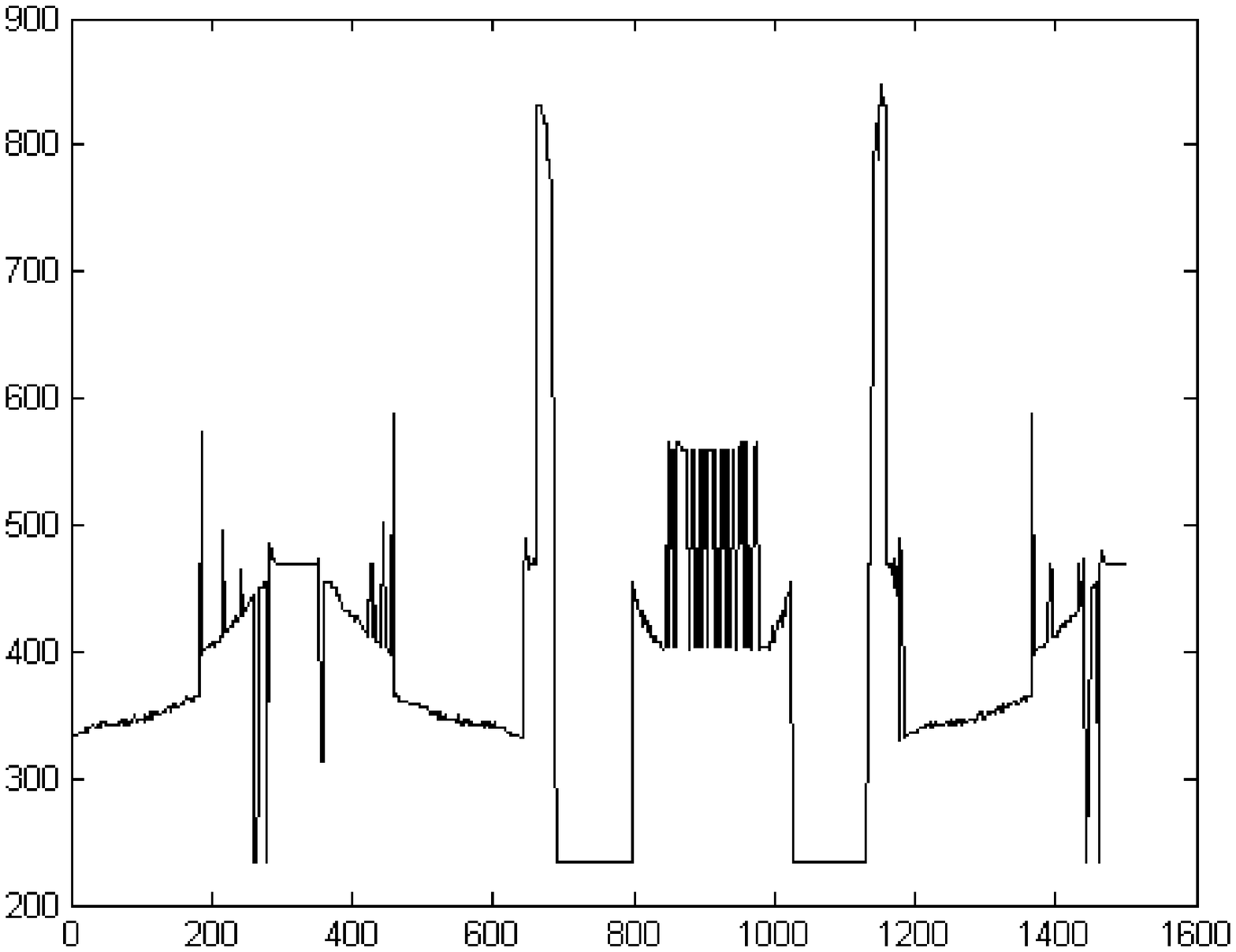,height=3.2cm,width=4.8cm} \\
\end{tabular}
\end{flushleft}
\footnotesize 
\begin{center}
Fig. 12 (a), (b), (c): Graphs of pitch for Do in Casio SA21 (Harmonica) by ASMDF, AMDF and Autocorrelation respectively.
\end{center}
\end{quotation}
\begin{quotation}
From the graphs we see that the noises and various encoding-decoding mismatches of signals are causing the distortions, but mainly the harmonica mode in the synthesizer here, being a well-known multiphonic device itself, has no fixed pitch value. Here too, ASMDF is much robust against mismatches and noises than those of autocorrelation and AMDF, which supports our claim.

\end{quotation}

\centering\mbox{\large{IV. ERROR ANALYSIS (of exp. no. 4)}}
\begin{quotation}
Let $P_c$ , $P_m$ and $P_s$ be the pitch contour found using autocorrelation, AMDF and ASMDF respectively calculated over the same window for the above three speeches of the two speakers. Let P be the true pitch value over the same window. Let us denote
\begin{center}
$e_s=(P-P_s)/P$ , $e_c=(P-P_c)/P$ and $e_m=(P-P_m)/P$
\end{center}
as the relative pitch error or percentage gross error. Let us define the standard deviation of the relative pitch error as
$$\sigma_e=\sqrt{\frac{1}{L_e-1}\sum_{i=1}^{L_e}e^2(i)-\overline{e}^2}$$
where $e(i)=e_s(i),e_m(i),e_c(i)$; $L_e$ being the length of each of the relative pitch error and $\overline{e}=\frac{1}{L_e}\sum_{i=1}^{L_e}e(i)$ is the mean pitch error. The experimental values of $\sigma_e$ viz. $\sigma_e(c)$, $\sigma_e(m)$ and $\sigma_e(s)$ for autocorrelation, AMDF and ASMDF respectively, is given in the following table:

\begin{center}
\epsfig{file=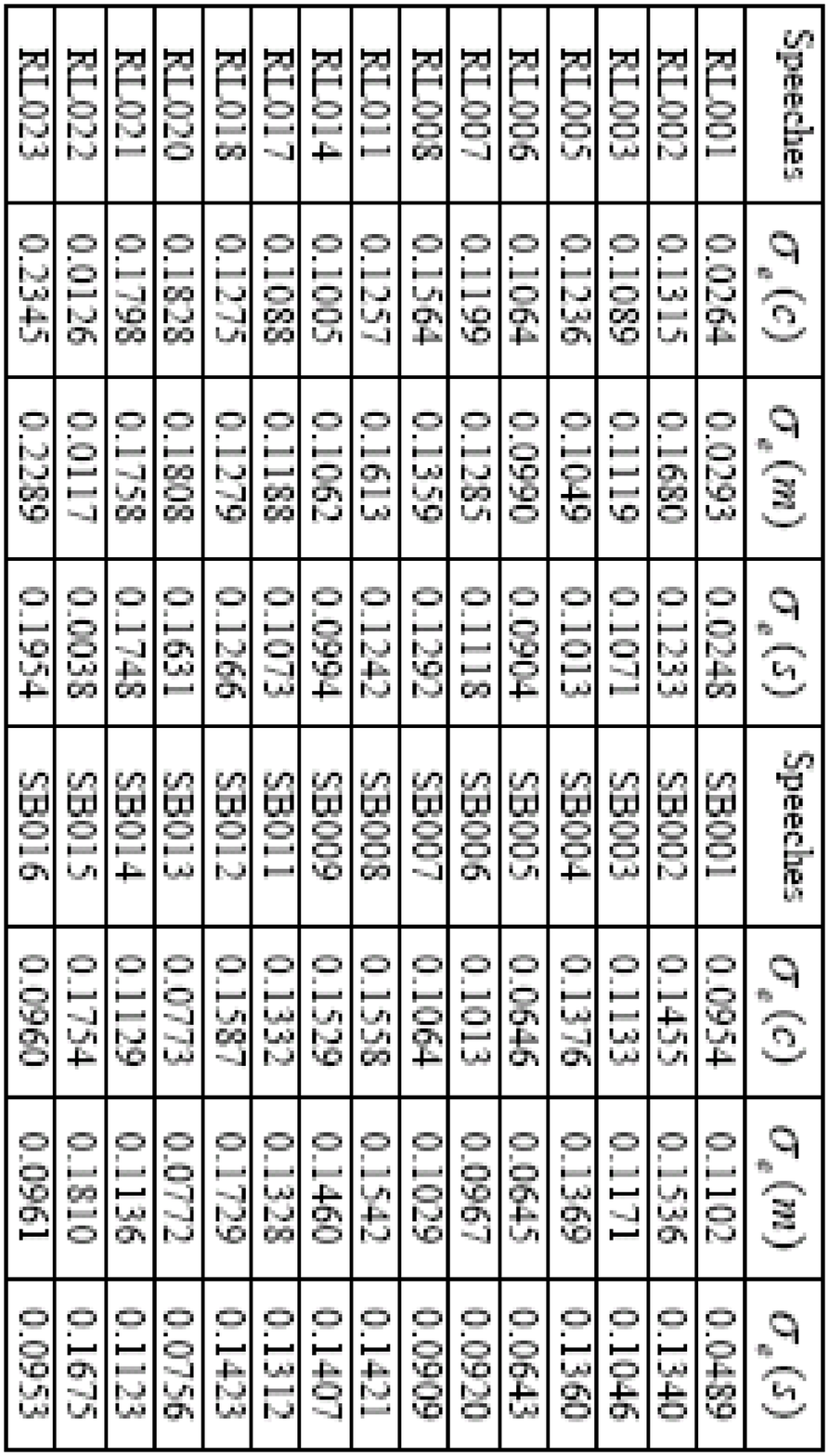,height=12cm,width=7cm,angle=90}
\end{center}

Here, ``$\sigma_e(s)$ for speech RL001 is 0.0248" means that standard deviation of gross pitch error for speech RL001 by ASMDF is 2.48\% with respect to the true pitch value. If the standard deviation of gross pitch error is more than 20\% (threshold) with respect to the true value, it's consistency is questionable. 

Also the correlation chart where correlation coefficients of $(P,P_s)$, $(P,P_c)$ and $(P,P_m)$ are calculated as $r_s$ , $r_c$ and $r_m$ respectively and are given as follows:

\begin{center}
\epsfig{file=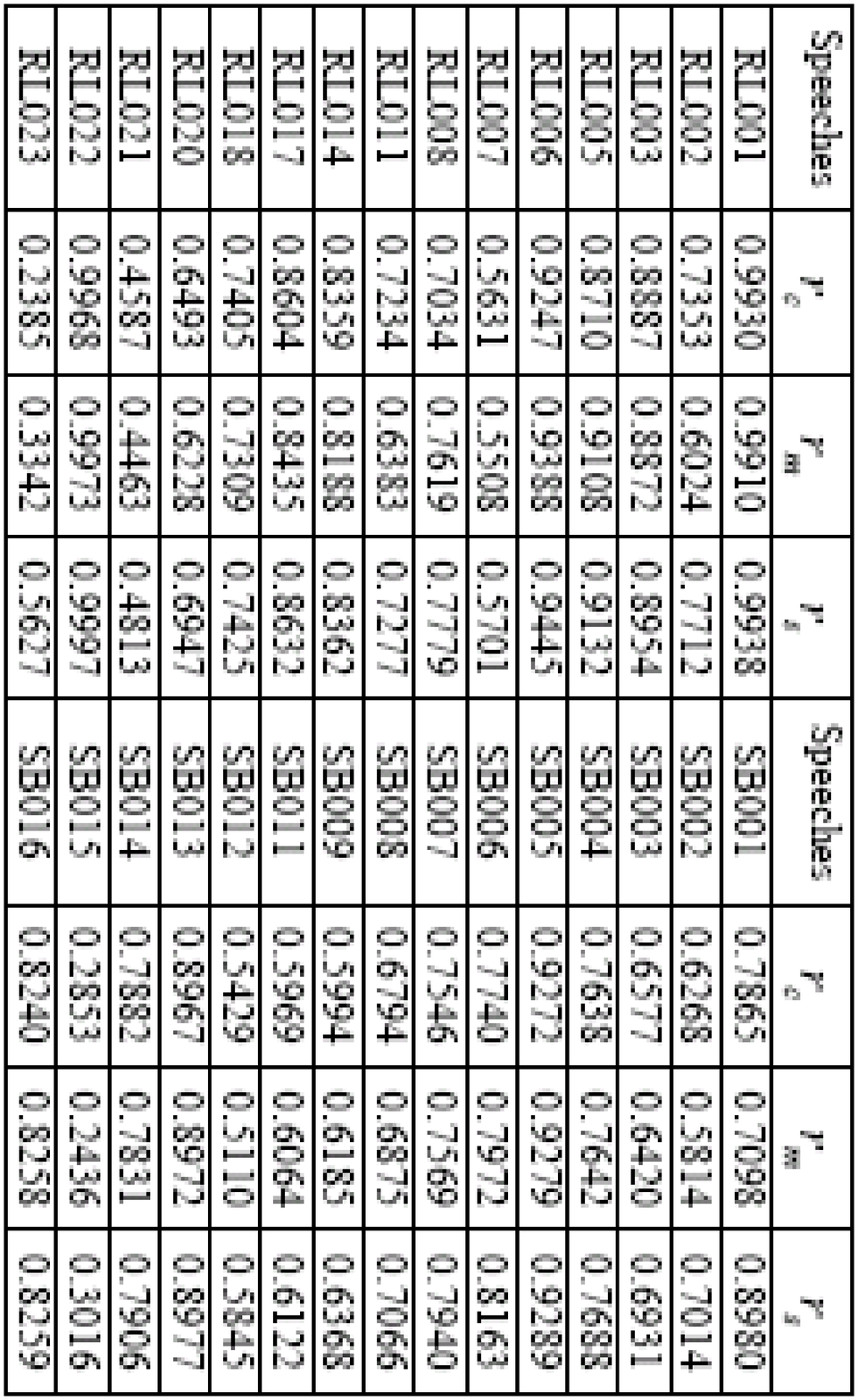,height=12cm,width=7cm,angle=90}
\end{center}
\end{quotation}
\centering\mbox{\large{V. CONCLUSION}}
\begin{quotation}
Based on the experimental results and error calculations it has been shown that ASMDF is useful in clean environment. It also makes a refinement of the autocorrelation-based methods to a great extent. The idea behind this method also leads to methods of extraction of other features of speech signals. Also there is a scope of using cutting and smoothing technique(s) for evaluation of pitch. The limitation of this method is the procedure to obtain the minimum values from the data set which are often used to fall behind the known limits viz. 80-200 Hz for males and 150-350 Hz for females.
\end{quotation}

\centering\mbox{\large{VI. APPENDIX}}

\centering\mbox{\normalsize{A. RELATIONSHIP BETWEEN ASMDF AND THE AUTOCORRELATION FUNCTION}}

\begin{quotation}

Note that, $g(k)$ can alternatively be defined as follows. Define the set
$$ C_k = \{ (i,j): \, 1 \leq i,j \leq n, \, i\neq j , \, |i-j| \mbox{ is divisible by } k \},
$$
for $k_0 \leq k \leq k_{max}$, where $|C_k|$ denotes the cardinality of $C_k$. Then 
$$
g(k) = \frac{1}{2|C_k|} \sum_{ (i,j) \in C_k}\, (y_i - y_j)^2,
$$
provided $C_k$ is non-empty. Otherwise $g$ is not defined. This follows from standard algebra with the formula for variance. Then it can be shown that 
\begin{eqnarray}
 E \{g(k)\}&=& \frac{1}{2|C_k|} \sum_{ (i,j) \in C_k}\, 2 \{ r_y(0) - r_y (|i-j|) \}\,  \, + \frac{1}{2|C_k|} \sum_{ (i,j) \in C_k}\, (Ey_i - Ey_j)^2\nonumber \\
 &\approx &r_y(0) \, \left\{1 - \lim_{N\rightarrow \infty} \,\frac{1}{N}\sum_{p=1}^{N}\left(1-\frac{p}{N}\right)\rho_y(pk)\right\}\, + \frac{1}{2|C_k|} \sum_{ (i,j) \in C_k}\, (Ey_i - Ey_j)^2,\nonumber
\end{eqnarray}
for large $n$ ($r_y$ and $\rho_y$ have already been defined in (2) and (3)). The bias term (second term) is being minimized in ASMDF based on $g(k)$. Note that the first term is added due to the noisy environment and is constant (equal to $r_y(0)$) for white noise error. Otherwise, even for well behaved stationary errors it is a slowly varying function ensuring robustness of our procedure under (1).

\end{quotation}\centering\mbox{\normalsize{B. COMPUTATIONAL COMPLEXITY}}
\begin{quotation}
Computational complexity of ASMDF (comes out roughly to be n(4n+5.5)), which is of order of $n^2$, is a bit higher than that of autocorrelation function (viz. 2n(n-1)) and AMDF (viz. (n-1)(3n-1)), which are also of orders of $n^2$, n being length of a window.
\end{quotation}
\centering\mbox{\textbf{REFERENCES}}

\small
\begin{quotation}
[1] The IViE Corpus. Phonetics Laboratory, University of Oxford and Department of Linguistics, University of Cambridge, 1997-2002.

[2] Paul Bagshaw, Fundamental Frequency Determination Algorithm (FDA) Evaluation Database. Centre for Speech Technology Research, University of Edinburgh, 1993.

[3] P. Cosi, S. Pasquin and E. Zovato. Auditory Modeling Techniques for Robust Pitch Extraction and Noise Reduction. Proc. of ICSLP (1998) paper 1053, 1998, http://citeseer.ist.psu.edu/cosi98auditory.html.

[4] David Gerhard. Pitch Extraction and Fundamental Frequency: History and Current Techniques. Technical Report TR-CS 2003-06: 1-22, November 2003, http://citeseer.ist.psu.edu/gerhard03pitch.html.

[5] Leah H. Jamieson, Goangshiuan S. Ying and Carl D. Michell. A Probabilistic Approach to AMDF Pitch Detection. Proceedings of the 1996 International Conference on Spoken Language Processing, Philadelphia, PA: 1201-1204, October 1996.

[6] Joseph P. Campbell Jr. Speaker Recognition: A Tutorial. Proceedings of the IEEE, 85(9): 1437-1462, September 1997.

[7] A. M. Goon, M. K. Gupta and B. Dasgupta. Fundamentals of Statistics Vol. 1. Sixth Edition. The World Press Private Ltd. July 1983.

[8] Benjamin Kedem. Spectral Analysis and Discrimination by Zero-crossings. Proceedings of the IEEE, 74(11): 1477-1493, November 1986.

[9] Sylvain Marchand. An Efficient Pitch-tracking Algorithm Using a Combination of Fourier Transforms. Proceedings of the COST G-6 Conference on Digital Audio Effects (DAFX-01), Limerick, Ireland: 170-174, December 2001, http://citeseer.ist.psu.edu/marchand01efficient.html.

[10] L. R. Rabiner. On the use of Autocorrelation Analysis for Pitch Detection. IEEE Trans. Acoust., Speech, Signal Processing, ASSP-25(1): 24-33, February 1977.

[11] L. R. Rabiner, M. J. Cheng, A. E. Rosenberg and C. A. McGonegal. A Comparative Performance Study of Several Pitch Detection Algorithms. IEEE Trans. Acoust., Speech, Signal Processing, ASSP-24(5): 399-417, October 1976.

[12] L. R. Rabiner, J. J. Dubnowski and R. W. Schafer. Real-Time Digital Hardware Pitch Detector. IEEE Trans. on Acoustics, Speech, and Signal Processing, ASSP-24(1): 2-8, February 1976.

[13] L. R. Rabiner, C. A. McGonegal and A. E. Rosenberg. A Semiautomatic Pitch Detector (SAPD). IEEE Trans. on Acoustics, Speech, and Signal Processing, ASSP-23(6): 570-574, December 1974.

[14] L. R. Rabiner and R. W. Schafer. System for Automatic Formant Analysis of Voiced Speech. Journal of the Acoustical Society of America, 47(2): 634-648, February 1970.

[15] M. J. Ross, H. L. Shaffer, A. Cohen, R. Freudberg and H. J. Manley. Average Magnitude Difference Function Pitch Extractor. IEEE Trans. Acoust., Speech, Signal Processing, ASSP-22: 353-362, October 1974.

[16] T. Shimamura and H. Kobayashi. Weighted Autocorrelation for Pitch Extraction of Noisy Speech. IEEE Trans. on Speech and Audio Processing, 9(7): 727-730, October 2001.

[17] L. Tan and M. Karnjanadecha. Pitch Detection Algorithm: Autocorrelation Method and AMDF. Proceedings of the 3rd International Symposium on Communications and Information Technology, 2: 551-556, September 2003.

[18] Kuo-Hwei Yuo, Tai-Hwei Hwang and Hsiao-Chuan Wang. Comparison of Autocorrelation-based Features and Projection Measure Technique for Speaker Identification. IEEE Trans. on Speech and Audio Processing, 13(4): 565-574, July 2005.

[19] R. D. Zilca, B. Kingsbury, J. Navratil and G. N. Ramaswamy. Pseudo Pitch Synchronous Analysis of Speech with Applications to Speaker Recognition. IEEE Trans. on Speech and Audio Processing, 14(2): 467-478, March 2006.

\end{quotation}

\end{document}